\documentclass[a4paper,11pt]{article}
\usepackage{jheppub}
\usepackage{amsmath,amssymb,amsthm,amsfonts,mathrsfs,mathtools}
\usepackage{tikz,commutative-diagrams}
\usepackage{graphicx,float,subfig,overpic}
\usepackage{bbm}

\def\bal#1\eal{\begin{align}#1\end{align}}

\title{Multiway junction conditions: Jackiw-Teitelboim gravity}
\author{Jia-Yin Shen}
\affiliation{Department of Physics, School of Physical Science and Technology, \\ Suzhou University of Science and Technology, Suzhou 215009, China}
\emailAdd{jiayin.shen@usts.edu.cn}
\abstract{A booklet is a geometric structure formed by gluing multiple bulk spacetimes along a common interface and imposing gravitational consistency conditions at the junction. We have systematically investigated the properties of booklet structures, constructed the booklet geometry, and derived the multiway junction conditions applicable at the interface. In this work, we provide a complete solution to the multiway junction conditions for booklets composed of bulks governed by JT gravity. By constructing invariants of the dilaton solution space, we classify all dilaton configurations into three inequivalent types, each exhibiting attractive, repulsive, or neutral behavior. Through continuous isometric transformations, each type is fixed to a standard form characterized by a single physical parameter, effectively eliminating redundant degrees of freedom. This process selects a distinguished class of Poincar\'e coordinates for each type. Expanding the constraint equations order by order breaks coordinate invariance beyond the leading and subleading orders. By jointly solving the junction and continuity conditions up to subleading order, we find that junctions are only allowed when a sufficient number of bulks carry attractive dilatons, as captured quantitatively by a equilibrium condition. We further analyze all possible combinations of different dilaton types and determine the shape of the interface along with the explicit form of the dilaton defined on it.
}
\arxivnumber{}

\begin{document}
\maketitle
\flushbottom

\section{Introduction\label{sec0}}
In a recent work~\cite{SPL,SPL1} of ours, we introduce a novel multi-boundary spacetime that is topologically distinct from the well-known replica wormholes---specifically, a structure where multiple bulk spacetimes are joined together along a common interface, which we refer to as a ``booklet.'' In this study, we undertake a meticulous exploration of the local geometry at the interface, developing a formalism termed ``booklet geometry,'' whose technical foundation is anchored in the ``reverse extension method.'' Through this methodology, we derive the ``multiway junction condition,'' which is applicable to the gluing of multiple spacetimes. This condition embodies the gravitational consistency of all spacetimes at the junction. Since no differential structure is defined at the junction of glued spacetimes (A detailed derivation can be found in~\cite{SPL1}), it is inherently impossible to directly define quantities such as the Einstein tensor. The reverse extension method, however, enables us to bypass this obstruction. Gravitational path integrals, by their nature, should sum over all possible spacetime topologies~\cite{Hawking1978quantum,carlip2001quantum,Almheiri:2019qdq,Penington:2019kki}, implying that the contributions from booklet structures must be incorporated. We anticipate that this innovative structure will offer fresh insights into the ongoing development of the black hole information paradox.
There have been interesting applications of the junction condition to holographic models~\cite{Liu:2025khw,Guo:2025sbm}. Recently, \cite{jiang2025GHZ} has explicitly presented solutions for booklet wormholes and revealed a duality between booklet wormholes and the Greenberger–Horne–Zeilinger state.

In the context of calculating the Page curve with replica wormholes, and given the pivotal role of JT gravity~\cite{Almheiri:2019qdq,Penington:2019kki}, this paper is devoted entirely to solving the multiway junction condition within the 2-dimensional JT gravity framework endowed with a negative cosmological constant. We have already undertaken extensive preparatory work in \cite{SPL1}, which includes: deriving the form of the multiway junction condition for general dilaton gravity using both the reverse extension method and by constructing the action and applying variational principles; as well as providing a preliminary solution for AdS spacetimes with tension at the interface in arbitrary dimensions. These results transition seamlessly to the JT gravity case. Notably, discussions of JT gravity often rely on a Weyl transformation to eliminate kinetic terms from the action~\cite{almheiri2015}. However, in~\cite{SPL1}, we observe that the actions before and after the Weyl transformation are not strictly equivalent, differing by a boundary term, which may influence the junction condition that serves as a boundary constraint. We rigorously prove the equivalence of the multiway junction condition under Weyl transformations.

In this paper, we do not glue all AdS$_2$ spacetimes directly at their conformal boundaries; rather, we introduce a cutoff curve at a finite distance near the conformal boundary. We place almost no constraints on the shape of the cutoff curve, thereby allowing it to approach the boundary in any arbitrary manner. In recent years, owing to the discovery of $T\bar{T}$ deformation theory~\cite{zamolodchikov2004,taylor2018tt,cardy2018tt,dubovsky2017,mcgough2018,cavaglia2016,kraus2018cutoff,Smirnov2017}, the holographic dictionary for gravity theories defined on finite patches has garnered increasing attention~\cite{Luca2020}. Imposing a finite cutoff in the bulk permits arbitrary shapes at the gluing interface, a feature that enables us to rigorously explore the physical implications encapsulated by the junction conditions. In fact, we find that the solution to the subleading order of constraints carries independent significance. However, in contrast to the approach in~\cite{Luca2020}, we demonstrate that it is unnecessary to compute to arbitrarily high orders, a claim that will be substantiated in this paper.

The paper is organized as follows: in section \ref{sec1}, we begin with a concise review of the construction of booklet geometry, the form of the multiway junction condition, and its application to dilaton gravity. Some of the conclusions presented will be directly applied within the paper. In section \ref{sec2}, first we solve the multiway junction condition for the extrinsic curvature up to subleading order within the framework of JT gravity, and provide a conceptual discussion of ``symmetry breaking induced by series expansion.'' Then we construct an invariant in the solution space of dilaton and use this invariant to classify dilatons. Relevant results involving the Lie algebra $\mathfrak{sl}(2,\mathbb R)$ will be derived in appendix \ref{appx2}. Furthermore, we jointly solve the continuity and junction conditions for dilaton up to subleading order. We will explore all possible scenarios in which spacetimes with different types of dilatons are glued together in varying proportions.

\section{Review of general results on multiway junction conditions\label{sec1}}

In our previous work~\cite{SPL,SPL1}, we derived multi-way junction conditions for Einstein gravity and dilaton gravity~\cite{BransDick1961,GRUMILLER2002327,Maeda2000,Turiaci2021,Henz2013}, which serve as physical consistency constraints at a common interface where an arbitrary number of bulk spacetimes meet. Based on these conditions, we constructed multi-way crossing geometries that we refer to as “booklet” geometries. In this section, we begin with a brief review of those results.

Consider $m+1$ bulk spacetimes (pages) glued together along a codimension-$1$ interface, $\Sigma$, locally forming a ``booklet'' $\mathscr P$. We distinguish different pages, denoted as $\mathscr V^{[i]},~i=0,1,\cdots,m$, by a ``page number'' $[i]$. The normal vector of $\mathscr V^{[i]}$ at $\Sigma$ is written as $n^{[i]}=\partial/\partial\ell^{[i]}$, where $\ell^{[i]}$ is the Gaussian coordinate along the normal direction in page $[i]$. For simplicity, we abbreviate $\ell^{[i]}$ to $\ell$ whenever the omitted page number can be easily infered from the context. The extrinsic curvature of $\Sigma$ within $\mathscr V^{[i]}$ is denoted by $K^{[i]}_{\bar\mu\bar\nu}$, where barred indices $\bar\mu,\bar\nu,\cdots$ refer to the intrinsic coordinate system on $\Sigma$, while unbarred indices $\mu,\nu,\cdots$ refer to coordinates inside a given page. Other physical quantities intrinsic to the junction are likewise always labelled by indices with bars. It should be noted that $x^\mu$ need not coincide with the Gaussian coordinate system $(x^{\bar\mu},\ell)$. The metrics $g^{[i]}_{\mu\nu}$ on all pages induce a common metric $h_{\bar\mu\bar\nu}$ on $\Sigma$. Without loss of generality, we assume that $h_{\bar\mu\bar\nu}$ is timelike and that all $n^{[i]}$ are spacelike normal vectors, so that a quantum field theory can reside within the interface. Of course, the derivation of the multiway junction conditions does not rely on this assumption; we make this assumption simply for the convenience of presentation. Then the multiway junction condition, e.g. for Einstein gravity, reads explicitly 
\begin{align}
    \label{multijc}
    \sum_{i=0}^m K^{[i]}_{\bar\mu\bar\nu}-K^{[i]}h_{\bar\mu\bar\nu}=-8\pi G_{\mathrm N}\bar S_{\bar\mu\bar\nu}.
\end{align}
where $\bar{S}_{\bar\mu\bar\nu}$ denotes the energy-momentum tensor localized on $\Sigma$. 

In the case of multi-way junctions in dilaton gravity, we start with a general action
\begin{align}
	\label{dilatongravity}
	I_{\mathscr V}=&\frac{1}{16\pi G_{\mathrm N}}\sum_{i=0}^m\int_{\mathscr V^{[i]}} \bigg\{(\Phi^{[i]})^2 R^{[i]}+\lambda\big(\partial\Phi^{[i]}\big)^2-\mathscr U^{[i]}\big((\Phi^{[i]})^2\big)\bigg\}\varepsilon^{[i]}-2\oint_\Sigma \bar\Phi^2K^{[i]}\bar\varepsilon_h,
\end{align}
where the dimensionless scalar field $(\Phi^{[i]})^2$ is the dilaton,  $2\bar\Phi^2K^{[i]}$ is the Gibbons-Hawking-York (GHY) boundary term~\cite{GH1977,Y1972}, and $\bar\Phi=\Phi^{[i]}|_\Sigma$ represents the boundary value of the dilaton. The $\mathscr U^{[i]}$ is the scalar potential of the dilaton, which has dimensions of Length$^{-2}$ and thus may depend on the characteristic length scale $L^{[i]}$ of $\mathscr V^{[i]}$ (for instance, if $\mathscr V^{[i]}$ is an AdS$_{d+1}$ spacetime, then $L^{[i]}$ would naturally be the AdS radius). $\varepsilon^{[i]}$ is the volume element of the page $\mathscr V^{[i]}$, while $\bar\varepsilon_h$ is that of $\Sigma$. Eq.\eqref{dilatongravity} is not the most general form of dilaton gravity, but it has the same form of 2D JT gravity, which is our primary focus in this paper. By taking into account the matter Lagrangian on the junction, $\bar{\mathcal L}_\text{matt}$, the form of the multiway junction conditions applicable to dilaton gravity can then be written as 
\begin{subequations}
	\label{dilatonjunction}
	\begin{align}
		\label{junctionpartPhi}
		&\sum_{i=0}^m 2\bar\Phi\, K^{[i]}+\lambda\partial_\ell\Phi^{[i]}=-4\pi G_{\mathrm N} \bar\varrho, \\
		\label{junctionpartK}
		&\sum_{i=0}^m \bar\Phi^2\,\big(K^{[i]}_{\bar\mu\bar\nu}-K^{[i]}\, h_{\bar\mu\bar\nu}\big)-2\bar\Phi \partial_\ell\Phi^{[i]}\, h_{\bar\mu\bar\nu}=-8\pi G_{\mathrm N} \bar{S}_{\bar\mu\bar\nu},
	\end{align}
 \end{subequations}
 where $\bar\varrho=-2\frac{\delta \bar{\mathcal L}_\text{matt}}{\delta\bar\Phi}$ is the strength of  the source on the interface that couples with the scalar. By a change of the frame of the action, namely if a Weyl transformation $g^{[i]}_{\mu\nu}\to (\Phi^{[i]})^{2\alpha}g^{[i]}_{\mu\nu},~ h_{\bar\mu\bar\nu}\to \bar\Phi^{2\alpha} h_{\bar\mu\bar\nu}$ of the metric is performed with the dimensionless constant $\alpha$ satisfying $\lambda+\alpha^2 d(d-1)-4\alpha d=0$, the kinetic term in the action can be eliminated, which is particularly useful in the calculation of JT gravity~\cite{Almheiri:2014cka}. As demonstrated in~\cite{SPL1}, the junction conditions across different frames are equivalent.~\footnote{Specifically, for $d=1$, after eliminating the kinetic term through a Weyl transformation, or equivalently setting $\lambda=0$, the form of the action remains unchanged.}

To simplify the discussion and reduce the technical complexity, we consider the leading approximation of the matter fields living on the interface $\Sigma$ whose sole effect is represented by a tension density $\chi$, and thus the action for the matter sector is given by $8\pi G_{\mathrm N}\,\bar{\mathcal L}_\text{matt}=-\bar\Phi^2\chi$. At this stage, the multiway junction conditions mean $\sum_{i=0}^mK^{[i]}_{\bar\mu\bar\nu}\propto \chi h_{\bar\mu\bar\nu}$. Building on this result, we will see that, in contrast to the classical gluing of two pages, the multi-page configuration introduces richer possibilities for the physics localized on the junction.

 Decomposing the extrinsic curvature $K^{[i]}_{\bar\mu\bar\nu}$ into its trace part $K^{[i]}$ and the symmetric traceless part $\xi^{[i]}$,
\begin{equation}
	\label{sumKi+chi=0}
	K^{[i]}_{\bar\mu\bar\nu}=\frac{1}{d}K^{[i]} h_{\bar\mu\bar\nu}+\xi^{[i]}_{\bar\mu\bar\nu},\qquad h^{\bar\mu\bar\nu}\xi^{[i]}_{\bar\mu\bar\nu}=0.
\end{equation}
The multiway junction conditions will imply $\sum_{i=0}^m \xi^{[i]}_{\bar\mu\bar\nu}=0$. We focus on the AdS$_{d+1}$ spacetime, which satisfies $R^{[i]}_{\ell\bar\sigma\bar\mu\bar\nu}=0$, and thus, using the Codazzi equation, we can derive the following constraint,
\begin{equation}
	\label{nablaKxi}
	\bar\nabla^{\bar\nu}\xi^{[i]}_{\bar\nu\bar\mu}=\bigg(1-\frac{1}{d}\bigg)\partial_{\bar\mu} K^{[i]}\ .
\end{equation}

The simplest solution has 
\bal
\xi^{[i]}_{\bar\mu\bar\nu}=0\,, \quad \forall i\,, 
\eal
therefore $\sum_{i=0}^m \xi^{[i]}_{\bar\mu\bar\nu}=0$ automatically  satisfies. Specifically, for $d>1$, Eq.\eqref{nablaKxi} implies that $K^{[i]}$ is constant. Combined with the Gauss equation,
\begin{equation}
 \label{GaussforAdS}
 R^{[i]}-2R^{[i]}_{\ell\ell}=\bar R+K^{[i]}_{\bar\mu\bar\nu}(K^{[i]})^{\bar\mu\bar\nu}-(K^{[i]})^2\,,
\end{equation}
the $K^{[i]}$ can be fully determineed, yielding a straightforward solution. 
In fact, for $d = 1$, $\xi^{[i]}_{\bar\mu\bar\nu}=0$ necessarily holds, as inferred from the assumption $h^{\bar\mu\bar\nu}\xi^{[i]}_{\bar\mu\bar\nu}=0$. Hence, the extrinsic curvature has only one degree of freedom, which is its trace. However, in this case, the Codazzi equation \eqref{nablaKxi} degenerates due to the factor $1-\frac{1}{d}$, and $K^{[i]}$ is not necessarily constant, leading to a rich variety of ways in which $\Sigma$ can be embedded in $\mathscr V^{[i]}$. This allows us to explore the physical implications of the multiway junction conditions while keeping the complexity of solving the equations relatively low.

In the next section, we will provide a detailed discussion of 2-dimensional JT gravity, which represents a significant special case of $d=1$ dilaton gravity. We assume that the tension on the interface is the only matter term. By setting $\lambda=0$ and $d=1$ in the dilaton gravity action, and abbreviating $(\Phi^{[i]})^2$ as $\phi^{[i]}$ and $\bar\Phi^2$ as $\bar\phi$, we obtain the multiway junction conditions
\begin{subequations}
	\label{JTjunction}
	\begin{align}
		\label{JTjunctionK}
		&\sum_{i=0}^mK^{[i]}=-\chi,\\
		\label{JTjunctionPhi}
		&\sum_{i=0}^m\partial_\ell\phi^{[i]}=-\bar\phi\chi, 
	\end{align}
\end{subequations}
which is applicable to JT gravity. Eqs.\eqref{JTjunctionK} and \eqref{JTjunctionPhi} are referred to as the junction conditions for the extrinsic curvature and the dilaton, respectively. The rest of this paper will be devoted to solving these two equations.

\section{Application: gluing multiple pages in JT gravity\label{sec2}}
The main purpose of this section is to apply the above general description to the theory of JT gravity. We will primarily consider a scenario where $m+1$ pages, each of which is a classical configuration of the JT gravity. 

JT gravity action is a special case of the general action~\eqref{dilatongravity} of dilaton gravity at $d=1$. The Weyl transformation $g^{[i]}_{\mu\nu}\to (\phi^{[i]})^{-\lambda/4}g^{[i]}_{\mu\nu}$ can be made to set  $\lambda=0$~\cite{almheiri2015}. 
Further choosing the potential to be linear in  $\phi^{[i]}$, we arrive at the action of JT gravity for each page. Adding the action of matter fields on a $d$-dimensional junciton we get
\begin{align}
	\label{JTaction}
	I_{\mathscr V}+I_{\text{matt}}=\frac{1}{16\pi G_{\mathrm N}}\sum_{i=0}^m\bigg\{\int_{\mathscr V^{[i]}}\phi^{[i]}\bigg(R^{[i]}+\frac{2}{(L^{[i]})^2}\bigg)\varepsilon^{[i]}-2\oint_\Sigma \bar\phi K^{[i]}\bar\varepsilon_h \bigg\}-\frac{1}{8\pi G_{\mathrm N}}\oint_\Sigma\bar\phi\chi\,\bar\varepsilon_h\,,
\end{align}
where we take the leading order approximaiton so that the details of the matter sector is encoded in the tension $\chi$ on the defect $\Sigma$, as we claimed in the previous section.
Vary the action with respect to the dilaton $\phi^{[i]}$  yields $R^{[i]}=-2/(L^{[i]})^2$, namely the geometry is simply AdS$_2$ on each page.

As reviewed in the previous section, the junction condition for dilaton gravity consists of the condition for the extrinsic curvature as well as the dilaton. In sections~\ref{curvature} and~\ref{dilaton} we discuss the two conditions respectively.

\subsection{Junction condition for the extrinsic curvature}\label{curvature}

In the following we consider gluing of multiple AdS$_2$ spacetime along a common junction that is close to a boundary of each AdS$_2$ page. 

The AdS$_2$ spacetime features two conformal boundaries. We can implement a cutoff in the AdS$_2$ spacetime by placng a curve $\Sigma$ near one of these boundaries and excluding the region between this boundary and $\Sigma$. The remaining  region from $\Sigma$ to the opposite boundary defines a page $\mathscr V^{[i]}$. The excluded region is denoted as $(\mathscr V^{[i]})^\complement$. It is evident that $\mathscr V^{[i]}$ and $(\mathscr V^{[i]})^\complement$ exhibit extrinsic curvatures with opposite signs at $\Sigma$. 
By aligning and identifying the cutoff curve on different pages to be the same $\Sigma$, we glue all $\mathscr V^{[i]}$ together to form a booklet $\mathscr P$. On each page, we consider the Poincaré patch and employ the following coordinates and metric on each page 
\begin{align}
	g^{[i]}=(L^{[i]})^2\cdot\frac{- \left(\mathrm dt^{[i]}\right)^2+ \left(\mathrm dz^{[i]}\right)^2}{\left(z^{[i]}\right)^2},\qquad z^{[i]}>0\ .
\end{align}
In addition, we choose an internal coordinate $u$ on the $1$-dimensional curve $\Sigma$, where the timelike vector $\partial_u$ points to the future and defines the orientation of $\Sigma$.  We can then express the shape of the cutoff curve $\Sigma$ in $\mathscr V^{[i]}$ as $(t,z)=\big(t^{[i]}(u),z^{[i]}(u)\big)$.
In the following we will sometimes drop the page label $[i]$ whenever it does not cause any confusion. In addition, since $u$ is an internal coordinate on $\Sigma$, it is shared by all pages and hence does not carry and page index $[i]$. 

We can reparameterize $u$ such that the metric $h$ satisfies the following condition~\cite{Maldacena:2016upp},
\begin{align}
	\label{hepsilon}
	h=-\frac{1}{\epsilon^2}\mathrm du^2,
\end{align}
where the small positive constant $\epsilon$ parameterizes the scale of the position of the cutoff surface near the boundary. Consequently, the normalized vector $\epsilon\partial_u$ tangent to $\Sigma$ can be represented as
\begin{align}
	\label{tangentvector}
	\epsilon\frac{\partial}{\partial u}=\epsilon\bigg(t^{[i]\prime}\frac{\partial}{\partial t}+ z^{[i]\prime}\frac{\partial}{\partial z}\bigg).
\end{align}
Using the fact that $g^{[i]}(\epsilon\partial_u,\epsilon\partial_u)=-1$, we can derive that
\begin{align}
	\label{SigmaParaEq}
	z^{[i]}(u)/L^{[i]}=\epsilon\sqrt{t^{[i]\prime}(u)^2-z^{[i]\prime}(u)^2}=\epsilon t^{[i]\prime}(u)+\mathcal O(\epsilon^3).
\end{align}
The logic behind expanding the series of the above expression is to treat the first equality in Eq.\eqref{SigmaParaEq} as a differential equation for $z^{[i]}(u)$, with $t^{[i]}(u)$ considered a known function. Consequently, the solution $z^{[i]}(u)$ depends on both $t^{[i]}(u)$ and $\epsilon$, specifically $z^{[i]}(u)=\mathscr F[t^{[i]}(u),\epsilon]$, the function $\mathscr F$ allowing us to express $z^{[i]}(u)$ as a series in $\epsilon$. It is clear that this series expansion assumes $t^{[i]}(u)$ and $\epsilon$ to be independent. While this assumption may seem obvious, it is crucial. Indeed, this assumption will be applied whenever we perform expansions of any quantities in the subsequent sections.

Continuing the iteration for Eq.\eqref{SigmaParaEq}, we can determine the higher-order terms of the series
\begin{align}
	\label{epsilonO3}
	z^{[i]}(u)/L^{[i]}\simeq \epsilon\sqrt{t^{[i]\prime}(u)^2-\big(\epsilon L^{[i]} t^{[i]\prime\prime}(u)\big)^2} =\epsilon t^{[i]\prime}(u)-\epsilon^3(L^{[i]})^2\frac{t^{[i]\prime\prime}(u)^2}{2t^{[i]\prime}(u)}
	+\mathcal O(\epsilon^5)\ .
\end{align}
Notice that the expansion of $z^{[i]}(u)$ include only odd powers of $\epsilon$. 

\subsubsection{Extrinsic curvature of the cutoff curve}

By definition, $g^{[i]}(n^{[i]},\partial_u)=0$, $g^{[i]}(n^{[i]},n^{[i]})=1$ and noting that the orientation of $\partial_u\wedge n^{[i]}$ is the same as that of $\partial_t\wedge\partial_z$, the normal vector $n^{[i]}$ can be represented as
\begin{align}
	\label{normalvecforAdS2}
	n^{[i]}=\epsilon\bigg(z^{[i]\prime}\frac{\partial}{\partial t}+t^{[i]\prime}\frac{\partial}{\partial z}\bigg)\ .
\end{align}
Now we calculate the extrinsic curvature $\mathcal K^{[i]}=K^{[i]}_{\mu\nu}\mathrm dx^\mu\mathrm dx^\nu$,
\begin{align}
	\label{K=Lnh/2}
	\notag
	 &\mathcal K^{[i]}=\frac{1}{2}\mathfrak L_{n^{[i]}} h=(L^{[i]})^2\bigg\{\frac{\mathrm dt^2-\mathrm dz^2}{z^3}\mathfrak L_{n^{[i]}} z+\frac{1}{z^2}\mathrm d(\mathfrak L_{n^{[i]}}z)\mathrm dz-\frac{1}{z^2}\mathrm d(\mathfrak L_{n^{[i]}} t)\mathrm dt\bigg\}\bigg|_{\mathscr T(\Sigma)} \\
	 &=-\bigg(\epsilon\frac{t^{[i]\prime}}{z^{[i]}}+\epsilon^3(L^{[i]})^2\frac{t^{[i]\prime\prime}z^{[i]\prime}-z^{[i]\prime\prime}t^{[i]\prime}}{(z^{[i]})^2}\bigg)h=-\bigg(\frac{1}{L^{[i]}}-\epsilon^2\mathsf{Sch}[t^{[i]}]L^{[i]}+\mathcal O(\epsilon^4)\bigg)h.
\end{align}
where $\mathfrak L_{n^{[i]}}$ denotes the Lie derivative along the direction $n^{[i]}$, and the symbol $\big|_{\mathscr T(\Sigma)}$ denotes restriction to the tangent space of $\Sigma$. 
Evidently, the expansion of $\mathcal K^{[i]}$ only contains even powers of $\epsilon$. Here, $\mathsf{Sch}[t^{[i]}]$ represents the Schwarzian derivative~\cite{Maldacena:2016hyu,Maldacena:2016upp} of $t^{[i]}$, 
\begin{align}
	\mathsf{Sch}[t^{[i]}](u)=\frac{2t^{[i]\prime}(u)t^{[i]\prime\prime\prime}(u)-3t^{[i]\prime\prime}(u)^2}{2t^{[i]\prime}(u)^2}.
\end{align}
and each order of the extrinsic curvature will depend on it or its derivative.
Consequently, the Schwarzian derivative of $ t^{[i]}$ possesses the dimensionality of Length$^{-2}$.

Although the symmetric and traceless part $\xi^{[i]}_{\bar\mu\bar\nu}$ of the extrinsic curvature vanishes, $K^{[i]}$ is not necesarrily a constant as we discussed at the end of section~\ref{sec1}. Explicitly, in addition to the leading order contribution $-\frac{1}{L^{[i]}}$, the higher order corrections $\mathcal O(\epsilon^2)$ are $u$-dependent. This property is  different from the $d>1$ case, due to the degeneracy of the Codazzi equation~\eqref{nablaKxi} when $d=1$ as we discussed above.

Before delving into the junction condition for the extrinsic curvature, let us turn our attention to the transformation between Poincar\'e coordinates $(t,z)$ and Gaussian coordinates $(u,\ell)$, which is convenient to compute the extrinsic curvature. The coordinates $(u,\ell)$ is defined in a vicinity of $\Sigma$ and $\partial_\ell$ is orthogonal to $\partial_u$ everywhere. We can extend the expressions \eqref{tangentvector} and \eqref{normalvecforAdS2} for the tangent vector $\epsilon\partial_u$ and the normal vector $n^{[i]}$ to the $(u,\ell)$ coordinate. The change of coordinate leads to the Jacobian matrix
\begin{align}
	\frac{\partial (t,z)}{\partial (u,\ell)}=
	\left(\begin{matrix}
			      \partial_ut(u,\ell) & \partial_uz(u,\ell)                     \\
			      \epsilon (u,\ell) \partial_uz(u,\ell) & \epsilon (u,\ell)   \partial_ut(u,\ell) \\
		\end{matrix}\right).
\end{align}
Note that in the above expression we have defined a function $\epsilon(u,\ell)$ by 
\bal
h_{uu}(u,\ell)=h(\partial_u,\partial_u)=-1/\epsilon(u,\ell)^2\,,\label{defeps}
\eal
at any point $(u,\ell)$. Since 
\bal
\partial_\ell h_{uu}=2K^{[i]}_{uu}\not=0\,, \label{dh}
\eal 
at the point of $\Sigma$, we observe that $\epsilon(u,\ell)$ depends on $\ell$; additionally, since $K^{[i]}_{uu}(u)$ is a function of $u$, $\epsilon(u,\ell)$ also depends on $u$. We will provide an expansion for $\epsilon (u,\ell)$ shortly. Thanks to the continuity of $z(u,\ell)$ and $\epsilon (u,\ell)$, the Jacobian determinant $\det\left|\frac{\partial (t,z)}{\partial(u,\ell)}\right|=(L^{[i]})^{-2}\frac{z(u,\ell)^2}{\epsilon(u,\ell)}\not=0$ holds within a sufficiently small open set, which means the coordinates $(u,\ell)$ are well-defined within this small open set. To get the explicit expression of the coordinate $(t,z)$ as a function of $(u,\ell)$, we adopt an iterative expansion along the normal $\ell$ direction. The first step is to express, e.g. $t(u,\ell),z(u,\ell)$, in terms of the $(t(u),z(u))$ and their derivatives on the junction, to $\mathcal O(\ell)$,
\begin{align}
	\label{tzul1}
	t(u,\ell)= t^{[i]}(u)+\epsilon z^{{[i]}\prime}(u)\ell+\mathcal O(\ell^2),\quad z(u,\ell)= z^{{[i]}}(u)+\epsilon t^{{[i]}\prime}(u)\ell+\mathcal O(\ell^2).
\end{align}
The use of the above equation allows us to expand $\epsilon(u,\ell)$,
\begin{align}
	\label{epsilonul}
	\epsilon(u,\ell)=\frac{z(u,\ell)/L^{[i]}}{\sqrt{[\partial_ut(u,\ell)]^2-[\partial_uz(u,\ell)]^2}}= \epsilon+\epsilon\ell/L^{[i]}-\epsilon^3L^{[i]}\mathsf{Sch}[t^{[i]}]\ell+\mathcal O(\epsilon^5)+\mathcal O(\ell^2).
\end{align}
We observe the appearance of the Schwarzian derivative $\mathsf{Sch}[t^{[i]}]$ which originates from the close relation between $\epsilon(u,\ell)$ and the extrinsic curvature $K^{[i]}_{uu}$~\eqref{defeps} and~\eqref{dh}. In fact,
\begin{align}
	K^{[i]}_{uu}=\frac{1}{2}\partial_\ell h_{uu}=-\frac{1}{2}\partial_\ell\epsilon(u,\ell)^{-2}\big|_{\ell=0}
	=\frac{1}{\epsilon^3}\partial_\ell\epsilon (u,\ell)\big|_{\ell=0}\simeq\frac{1}{\epsilon^2L^{[i]}}-\mathsf{Sch}[t^{[i]}]L^{[i]}.
\end{align}
To get the relation between $(t,z)$ and $(u,\ell)$ deeper in the bulk, we iterate the above expansion to get the expression of  $t(u,2\ell)$ and $z(u,2\ell)$. In practice, we write a similar expansion as~\eqref{epsilonul} but replace right-hand-side(RHS) by the functions evaluated at depth $\ell$, namely replacing $t^{[i]}(u)$,  $z^{[i]}(u)$ and $\epsilon$ by $t(u,\ell)$, $z(u,\ell)$, and $\epsilon(u,\ell)$ respectively in the RHS Eq.\eqref{tzul1}, we derive a high-order expansion for $t(u,2\ell)$. By making the substitution $2\ell\to\ell$ so that the expansion takes a conventional form, we arrive at
\begin{align}
	\notag
	 & t(u,\ell)=t^{[i]}(u)+\epsilon z^{[i]\prime}(u)\ell+\frac{1}{4}\big(\epsilon z^{[i]\prime}(u)/L^{[i]}+\epsilon^2 t^{[i]\prime\prime}(u)\big)\ell^2+\mathcal O(\ell^3)+\mathcal O(\epsilon^3),\\
	 & z(u,\ell)= z^{[i]}(u)+\epsilon t^{[i]\prime}(u)\ell+\frac{1}{4}\big(\epsilon t^{[i]\prime}(u)/L^{[i]}+\epsilon^2z^{[i]\prime\prime}(u)\big)\ell^2+\mathcal O(\ell^3)+\mathcal O(\epsilon^3),
\end{align}
where the term $\frac{1}{4}\epsilon^2z^{[i]\prime\prime}(u)\ell^2$ is actually of $\mathcal O(\epsilon^3)$, and thus can be neglected, while the other terms are all of order $\epsilon$. In addition, this result agrees with~\eqref{epsilonul} upto the $\mathcal{O}(\ell)$ order as expected. Notice that we can also reverse this analysis and solve for the transformation $u(t,z)$ and $\ell(t,z)$ in terms of $(t,z)$. More generally, it is possible to choose other different coordinates for the (following) computations. As shown here the concrete expression in different coordinates might be diferent, but the procedure works exactly as we will present in the later section. We further discussed the change of coordinates in appendix~\ref{changecoord}.  

\subsubsection{Near-boundary expansion of the junction condition for the  curvature\label{subsec2B}}
Returning to the discussion of the junction condition, we have already obtained the result \eqref{JTjunction}, which is applicable to JT gravity.
In this section, we focus on Eq.\eqref{JTjunctionK}, which directly yields
\begin{align}
	\label{seriesjunctionK}
	\chi=-\sum_{i=0}^mK^{[i]}=\sum_{i=0}^m\frac{1}{L^{[i]}}+\epsilon^2\sum_{i=0}^mL^{[i]}\mathsf{Sch}[t^{[i]}]+\mathcal O(\epsilon^4).
\end{align}
Matching terms of the same order on both sides, we have
\begin{align}
	\label{AdS2tension}
	\chi=\frac{1}{L^{[0]}}+\frac{1}{L^{[1]}}+\cdots+\frac{1}{L^{[m]}},
\end{align}
and
\begin{align}
	\label{sumSchLi}
	\sum_{i=0}^mL^{[i]}\mathsf{Sch}[t^{[i]}]=0.
\end{align}
At the leading order, we observe that the junction condition for the extrinsic curvature does not impose any restrictions on the embedding of $\Sigma$ into $\mathscr V^{[i]}$ since there are no terms containing $t^{[i]}$. However, at $\mathcal O(\epsilon^2)$, it does constrain the embedding of $\Sigma$ to satisfy Eq.\eqref{sumSchLi}.

Let us clarify one point. When dealing with the junction condition between two pages, the continuity condition $n^{[0]}+n^{[1]}=0$ 
allows us to establish a shared coordinate $(t,z)$ independent of indice $[i]$. However, in the case of the $m+1$ pages currently under discussion, 
we must select separate sets of coordinates within each $\mathscr V^{[i]}$. Fortunately, the junction condition remains independent of the choice. To emphasize this, we derived the junction condition in the previous paper \cite{SPL1} using an explicitly geometrically invariant approach.

\subsection{Junction condition for the dilaton}\label{dilaton}

\subsubsection{Classification of the dilaton profiles\label{sec3}}
The equation of motion of the dilaton field in JT gravity reads
\begin{align}
	\label{motiondilaton}
	\nabla^{[i]}_\mu\partial_\nu\phi^{[i]}-g^{[i]}_{\mu\nu}\bigg(\Box^{[i]}\phi^{[i]}-\frac{1}{(L^{[i]})^2}\phi^{[i]}\bigg)=0\ .
\end{align}
As demonstrated explicitly in appendix~\ref{appx1}; the solution can be parameterized as
\begin{align}
	\label{solutionofdilaton}
	\phi^{[i]}(t,z)=A\cdot\frac{t}{z}+B\cdot\frac{t^2-z^2}{z}+C\cdot\frac{1}{z}.
\end{align}
The coefficients $A, B, C$ can take on any real values, where $A$ is dimensionless; $B$ has the dimensionality of Length$^{-1}$; and $C$ of Length$^1$. Since these three parameters are arbitrary, and we focus solely on the  dilaton within a certain $\mathscr V^{[i]}$ in this section, without considering the interrelations among the dilatons of different pages, there is no need to distinguish these parameters with a page number $[i]$. Consequently, all possible $\phi^{[i]}$ form a vector space, denoted as $\mathsf{Sol}(\phi)$. We have selected a specific set of basis vectors for this space, denoted as $\vartheta_A,\vartheta_B,\vartheta_C\in\mathsf{Sol}(\phi)$,
\begin{align}
	\vartheta_A =\frac{t}{z},\quad \vartheta_B=\frac{t^2-z^2}{z},\quad  \vartheta_C=\frac{1}{z},
\end{align}
and we shall compute all the Killing vectors $\xi$ for the AdS$_2$ metric defined by  $\mathfrak L_{\xi}g^{[i]}=0$. The resulting Killing vector $\xi$ is a linear combination of 3 basis vectors $\xi=a\xi_a+b\xi_b+c\xi_c$
\begin{align}
	\label{generatorofsl2R}
	\xi_a=t\frac{\partial}{\partial t}+z\frac{\partial}{\partial z},\quad
	\xi_b=(t^2+z^2)\frac{\partial}{\partial t}+2tz\frac{\partial}{\partial z},\quad
	\xi_c=\frac{\partial}{\partial t}\,,
\end{align}
where the coefficients $a, b, c$ are real. 
The Lie algebra of the Killing vectors reads
\begin{align}
	\label{liestructure}
	[\xi_a,\xi_b]=\xi_b,\quad [\xi_a,\xi_c]=-\xi_c,\quad [\xi_b,\xi_c]=-2\xi_a\,, 
\end{align}
which is isomorphic to the Lie algebra $\mathfrak{sl}(2,\mathbb R)$. Therefore $\mathsf{Sol}(\phi)$ also constitutes a representation space of $\mathfrak{sl}(2,\mathbb R)$, and we have
\begin{align}
	\notag
	 & \xi_a(\vartheta_A)=0, \quad \xi_a(\vartheta_B)=\vartheta_B,\quad\xi_a( \vartheta_C)=-\vartheta_C;  \\
	\notag
	 & \xi_b(\vartheta_A)=-\vartheta_B,\quad\xi_b(\vartheta_B)=0,\quad\xi_b(\vartheta_C)=-2\vartheta_A; \\
	 & \xi_c(\vartheta_A)=\vartheta_C,\quad\xi_c(\vartheta_B)=2\vartheta_A,\quad \xi_c(\vartheta_C)=0\ . \label{thetarep}
\end{align}
Comparing~\eqref{liestructure} with~\eqref{thetarep}, it is clear to see that $\mathfrak{sl}(2,\mathbb R)$ and $\mathsf{Sol}(\phi)$ are isomorphic as $\mathfrak{sl}(2,\mathbb R)$ representation, with the isomorphism mapping as $\sigma:\xi_a\to  \vartheta_A,\xi_b\to\vartheta_B,\xi_c\to\vartheta_C$.  We can use this isomorphism to identify invariants within $\mathsf{Sol}(\phi)$.

The integral curves of any Killing vector $\xi=a\xi_a+b\xi_b+c\xi_c$ will give rise to a one-parameter family of local diffeomorphism within an open set, denoted as $\exp(s\xi)$, where $s$ serves as a parameter. For any point $q$ in spacetime, $\exp(s\xi)(q)$ represents another point obtained by Lie transport of $q$ along the integral curve of the vector field $\xi$ over a distance parameterized by $s$. Let the coordinates of the initial point $q$ be $(t,z)$, and the coordinates of the endpoint $\exp(s\xi)(q)$ be denoted as $\big(t(s),z(s)\big)$. The two coordinates can be related through the pullback map associated with the diffeomorphism $\exp(s\xi)$, that is, $\exp(s\xi)^*(t,z) = \big(t(s),z(s)\big)$, with $\big(t(0),z(0)\big)=(t,z)$. 
It is possible to derive the expression~\eqref{tskappat} for $\exp(s\xi)^*$ corresponding to general $\xi$, aligning with the widely adopted form found in other references~\cite{grumiller2021, mertens2023, sarosi2017, Luca2020}---see appendix \ref{appx2}. But for later convenience, we directly integrate to get the continuous transformations associated with $\xi_a, \xi_b, \xi_c$,
\begin{align}
	\label{oneparatransmation}
	\notag
	 & \exp\big(a\xi_a\big)^*(t,z)=\big(e^at,e^az\big);
	\qquad \exp\big(c\xi_c\big)^*(t,z)=\big(t+c,z\big); \\
	 & \exp\big(b\xi_b\big)^*(t,z)=\bigg(\frac{t-b(t^2-z^2)}{1-2bt+b^2(t^2-z^2)},\frac{z}{1-2bt+b^2(t^2-z^2)}\bigg)\ .
\end{align}
The action of these three special transformations on $\mathsf{Sol}(\phi)$ are
\begin{align}
	\notag
	 & \exp(a\xi_a)^*\vartheta_A=\vartheta_A,\quad\exp(a\xi_a)^*\vartheta_B =e^a\vartheta_B,\quad\exp(a\xi_a)^*\vartheta_C=e^{-a}\vartheta_C\\
	\notag
	 & \exp(b\xi_b)^*\vartheta_A=\vartheta_A-b\vartheta_B,\quad\exp(b\xi_b)^* \vartheta_B=\vartheta_B,\quad\exp(b\xi_b)^*\vartheta_C=\vartheta_C-2b\vartheta_A +b^2\vartheta_B \\
	 & \exp(c\xi_c)^*\vartheta_A=\vartheta_A +c\vartheta_C,\quad \exp(c\xi_c)^* \vartheta_B=\vartheta_B+2c\vartheta_A +c^2\vartheta_C,\quad \exp(c\xi_c)^* \vartheta_C=\vartheta_C.
\end{align}
Using the equations above, we can readily eliminate the majority of redundant degrees of freedom in $\phi^{[i]}(t,z)=A\vartheta_A+B\vartheta_B+C\vartheta_C$. Now, let us assume $A^2-4BC>0$. In this scenario, if $B=C=0$, we refrain from making any transformations and maintain the form of $\phi^{[i]}=A\frac{t}{z}$ unaltered. Consequently, we assume that $C\not=0$, and under this condition, we can apply continuous transformations generated by $\xi_b,\xi_c$,
\begin{align}
	\exp\bigg(\mp\frac{C}{\sqrt{A^2-4BC}}\xi_c\bigg)^*\exp\bigg(\frac{A\mp\sqrt{A^2-4BC}}{2C}\xi_b\bigg)^*\phi^{[i]}(t,z)=\pm\sqrt{A^2-4BC}\frac{t}{z}.
\end{align}
Next, we can redefine the parameter as $\widetilde A=\pm\sqrt{A^2-4BC}$. In other words, through a metric-preserving transformation, we have transformed $\phi^{[i]}$ into the form $\widetilde A\frac{t}{z}$. However, if $C=0$, we can assume $B\not=0$,  then the following transformation can be applied,
\begin{align}
	\exp\bigg(\pm\frac{B}{\sqrt{A^2-4BC}}\xi_b\bigg)^*\exp\bigg(\frac{-A\pm\sqrt{A^2-4BC}}{2B}\xi_c\bigg)^*\phi^{[i]}(t,z)=\pm\sqrt{A^2-4BC}\frac{t}{z}.
\end{align}
By redefining the parameter $\widetilde A$, it can be observed that $\phi^{[i]}$ can still be expressed in the form of $\widetilde A\frac{t}{z}$. Note that at this point, either $\widetilde B$ or $\widetilde C$  equals 0. Before and after the coordinate transformation, we have $\widetilde A^2-4\widetilde B\widetilde C=A^2-4BC$. We can see that the value $A^2-4BC$ remains unchanged. An interesting fact to note is that if we consider the reverse process of the aforementioned transformation, such as $A\frac{t}{z}\Longrightarrow 3A\frac{t}{z}+2A\frac{t^2-z^2}{z}+A\frac{1}{z}\Longrightarrow -A\frac{t}{z}$, it is shown that $A\frac{t}{z}$ and $-A\frac{t}{z}$ can also be transformed into each other. Therefore, the dilaton with $A^2-4BC > 0$ can always be brought into the following form through a coordinate transformation,
\begin{align}
	\label{phiA}
	\phi^{[i]}=A\dfrac{t}{z},\quad A>0.
\end{align}

Now, consider the case where $A^2-4BC\leqslant 0$. Excluding the trivial solution $(A,B,C)=(0,0,0)$, then there must be either $B\not=0$  or $C\not=0$, and for these two cases, we can make respective transformations as
\begin{align}
	\notag
	 & \exp\bigg(-\frac{A}{2B}\xi_c\bigg)^*\phi^{[i]}(t,z)=B\cdot\frac{t^2-z^2}{z}+\bigg(C-\frac{A^2}{4B}\bigg)\frac{1}{z}, \\
	 & \exp\bigg(\frac{A}{2C}\xi_b\bigg)^*\phi^{[i]}(t,z)=\bigg(B-\frac{A^2}{4C}\bigg)\frac{t^2-z^2}{z}+C\cdot\frac{1}{z},
\end{align}
by redefining the parameters $\widetilde B$ and $\widetilde C$, we find that $\phi^{[i]}$ can inevitably be transformed into the form
\begin{align}
	\label{phiBC}
	\phi^{[i]}(t,z)=\widetilde B\cdot\frac{t^2-z^2}{z}+\widetilde C\cdot\frac{1}{z},
\end{align}
where $\widetilde A=0$, we still have $A^2-4BC=\widetilde A^2-4\widetilde B\widetilde C$.  As a result, $\widetilde B\widetilde C\geqslant 0$. First consider the case where $\widetilde B\widetilde C>0$, namely $A^2-4BC<0$. Then, it is necessary that both $\widetilde B\not=0$ and $\widetilde C\not=0$ simultaneously, while $\widetilde B$ and $\widetilde C$ sharing the same sign. We can perform the following transformations,
\begin{align}
	\exp\bigg(\frac{1}{2}\ln\frac{\widetilde C}{\widetilde  B(L^{[i]})^2}\xi_a\bigg)^*	\bigg(\widetilde  B\cdot\frac{t^2-z^2}{z}+\widetilde  C\cdot\frac{1}{z}\bigg)=\pm\frac{\sqrt{\widetilde  B\widetilde  C}}{L^{[i]}}\bigg(\frac{t^2-z^2}{z}+(L^{[i]})^2\frac{1}{z}\bigg),
\end{align}
when $\widetilde B,\widetilde C>0$, the expression above is positive; conversely, when $\widetilde B,\widetilde C<0$, it is negative.
Once again, by redefining the parameter $\frac{\sqrt{\widetilde  B\widetilde  C}}{L^{[i]}}\to B$, the dilaton field is transformed into the form
\begin{align}
	\label{phiB}
	\phi^{[i]}(t,z)=\pm B\bigg(\frac{t^2-z^2}{z}+(L^{[i]})^2\frac{1}{z}\bigg), \quad B>0,
\end{align}
where there are no transformations to connect the two cases with $+$ and $-$.

Finally, for Eq.\eqref{phiBC}, when $A^2-4BC=-4\widetilde B\widetilde C=0$, it is inevitable that either $\widetilde B=0$ or $\widetilde C=0$, excluding the case of $\widetilde B=\widetilde C=0$ which corresponds to the trivial solution $\phi^{[i]}=0$. In this scenario, $\phi^{[i]}(t,z)$ is either proportional to $\frac{t^2-z^2}{z}$ or proportional to $\frac{1}{z}$. However, these two solutions can also be connected by a metric-preserving transformation,
\begin{align}
	\exp\bigg(\frac{1}{c}\xi_b\bigg)^*\exp\big(c\xi_c\big)^*\bigg(\frac{t^2-z^2}{z}\bigg)=c^2\cdot\frac{1}{z}.
\end{align}
Therefore, $\phi^{[i]}$ can always be transformed into the form $\widetilde C\cdot\frac{1}{z}$. Furthermore, through the transformation $\exp\bigg(\ln\frac{|\widetilde C|}{L^{[i]}}\cdot\xi_a\bigg)$, it can be proven that $\phi^{[i]}$ with $A^2-4BC=0$ can always be transformed into the form
\begin{align}
	\label{phiC}
	\phi^{[i]}(t,z)=\pm L^{[i]}\cdot\frac{1}{z}.
\end{align}
Similar to Eq.\eqref{phiB}, in the above equation, the two cases of taking $+$ and $-$ cannot be transformed into each other.

\begin{figure}[htbp!]
	\centering
	\subfloat[An equivalent class of Type $+$ with $A^2-4BC$ as a positive constant]{\includegraphics[width=0.3\columnwidth]{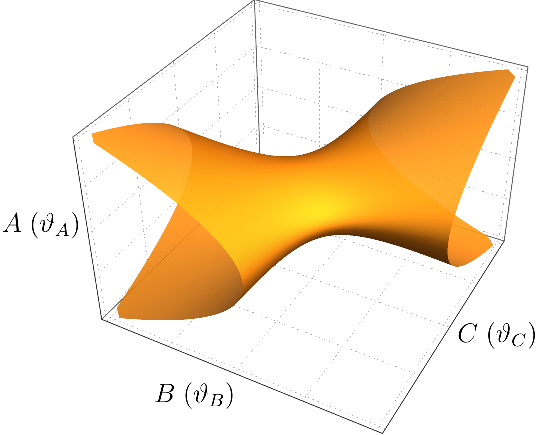}}\hspace{2cm}
	\subfloat[Dilatons of Type $0$ with $A^2-4BC=0$ and the trivial solution $(0,0,0)$ (black dot)]{\includegraphics[width=0.3\columnwidth]{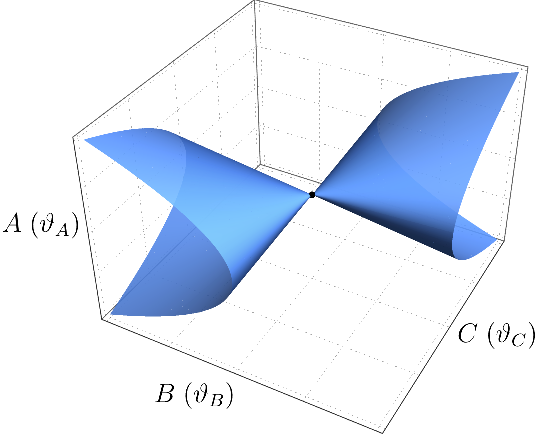}}\\
	\subfloat[Dilatons of Type $-$ with $A^2-4BC<0$]{\includegraphics[width=0.3\columnwidth]{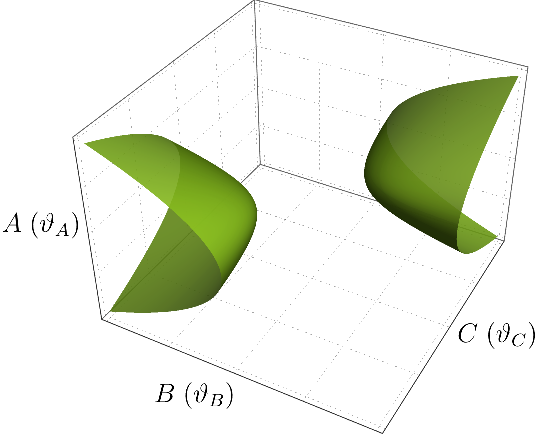}}\hspace{2cm}
	\subfloat[The solution space of all dilatons]{\includegraphics[width=0.3\columnwidth]{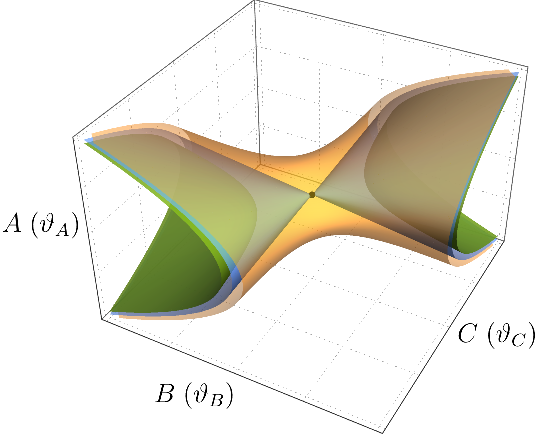}}
	\caption{Dilatons are classified based on the values of $A^2-4BC$, with Type $0$ and Type $-$ further divided into two disconnected parts. The topological disconnectedness of Type $0$ is due to the trivial solution $(0,0,0)$, which  is inequivalent to all non-trivial dilatons\label{DilatonSpace}}
\end{figure}

Summarizing the above results, for two dilaton fields $\phi,\tilde\phi\in\mathsf{Sol}(\phi)$, if they can be transformed into each other by a continuous metric-preserving transformation $\exp(\xi)\in\mathrm{SL}(2,\mathbb R)$, such that $\tilde\phi=\exp(\xi)^*\phi$, we refer to these two dilatons as equivalent, denoted as $\tilde\phi\sim\phi$. It is easy to prove that this is a mathematical equivalence relation, and all equivalent dilatons form an equivalence class. Since $\exp(\xi)^*$ is a reversible mapping, the trivial dilaton $\phi=0$ cannot be equivalent to non-trivial solutions. From the calculations above, we observe that all elements within an equivalence class share the same value of $A^2-4BC$. If two equivalence classes have the same non-positive $A^2-4BC$, they can be further distinguished by the sign of $B$ or $C$. Conversely, if two dilatons have different $A^2-4BC$, they must belong to different equivalence classes. Therefore, we can classify all non-trivial dilatons according to the sign of $A^2-4BC$. We depict the topological structure of the dilaton solution space in figure \ref{DilatonSpace}.

In fact, $A^2-4BC$ can be considered as a characteristic of the equivalence classes, and we will prove that $A^2-4BC$ corresponds to an invariant of the Lie algebra. Through the isomorphism mapping $\sigma$, it becomes an invariant of $\mathsf{Sol}(\phi)$. To show this, we first construct a function that, for given $\phi^{[i]}$, can extract the corresponding $A^2-4BC$. Utilizing the dual basis $\vartheta^{*A},\vartheta^{*B},\vartheta^{*C}$ with respect to $\vartheta_A , \vartheta_B , \vartheta_C$,  satisfying $\vartheta^{*\alpha}(\vartheta_\beta)=\delta^\alpha_\beta, \quad\forall \alpha,\beta\in\big\{A,B,C\big\}$, we define an inner product $\Omega$ for the $3$D vector space $\mathsf{Sol}(\phi)$,
\begin{align}
	\Omega=\frac{1}{2}\vartheta^{*A}\otimes \vartheta^{*A}-\vartheta^{*B}\otimes \vartheta^{*C}-\vartheta^{*C}\otimes \vartheta^{*B},
\end{align}
then, for any $\phi^{[i]}=A\vartheta_A+B \vartheta_B +C \vartheta_C$, $\Omega$ yields
\begin{align}
	\Omega\big(\phi^{[i]},\phi^{[i]}\big)=\frac{A^2-4BC}{2},
\end{align}
which is precisely what we desire. By using the transpose $\sigma^{\mathsf t}:\vartheta^{*A}\to \xi^{*a}, \vartheta^{*B}\to \xi^{*b},\vartheta^{*C}\to \xi^{*c}$ of the isomorphism $\sigma$, where the dual basis $\xi^{*\alpha}$ satisfies $\xi^{*\alpha}(\xi_\beta)=\delta^\alpha_\beta$ for $\alpha,\beta\in\big\{a,b,c\big\}$, we can obtain a bilinear form $\mathfrak K$ defined on $\mathfrak{sl}(2,\mathbb R)$,
\begin{align}
	\mathfrak K\coloneqq \sigma^{\mathsf t}\big(\Omega\big)=\frac{1}{2}\xi^{*a}\otimes\xi^{*a}-\xi^{*b}\otimes\xi^{*c}-\xi^{*c}\otimes\xi^{*b}.
\end{align}
In fact, $4\mathfrak K$ is essentially the Killing form (or the Casimir element) of the Lie algebra, thus $\mathfrak K\big([\xi,\xi_1],\xi_2\big)+\mathfrak K\big(\xi_1,[\xi,\xi_2]\big)=0$ holds for any $\xi,\xi_1,\xi_2\in\mathfrak{sl}(2,\mathbb R)$. Then, by taking $\xi_1=\xi_2$, and considering $\sigma(\xi_1)=\phi^{[i]}$, we obtain
\begin{align}
	0=2\mathfrak K\big([\xi,\xi_1],\xi_1\big)=2\Omega\big(\xi\phi^{[i]},\phi^{[i]}\big)=\frac{\mathrm d}{\mathrm ds}\Omega\big(\exp(s\xi)^*\phi^{[i]},\exp(s\xi)^*\phi^{[i]}\big)\big|_{s=0}.
\end{align}
Due to the arbitrariness of $\phi^{[i]}$, by selecting a fixed parameter $s_0$, we can replace $\phi^{[i]}$ with $\exp(s_0\xi)^*\phi^{[i]}$, and the above equation still holds, namely:
\begin{align}
	\notag
	0&=\frac{\mathrm d}{\mathrm ds}\Omega\big(\exp((s+s_0)\xi)^*\phi^{[i]},\exp((s+s_0)\xi)^*\phi^{[i]}\big)\big|_{s=0}\\
	&=\frac{\mathrm d}{\mathrm ds}\Omega\big(\exp(s\xi)^*\phi^{[i]},\exp(s\xi)^*\phi^{[i]}\big)\big|_{s=s_0}.
\end{align}
Integrating the above equation along the integral curve of the vector field $\xi$,
\begin{align}
	\notag
	0&=\int_0^1\frac{\mathrm d}{\mathrm ds}\Omega\big(\exp(s\xi)^*\phi^{[i]},\exp(s\xi)^*\phi^{[i]}\big)\,\mathrm ds=\int_0^1\,\mathrm d\Omega\big(\exp(s\xi)^*\phi^{[i]},\exp(s\xi)^*\phi^{[i]}\big)\\
	&=\Omega\big(\exp(\xi)^*\phi^{[i]},\exp(\xi)^*\phi^{[i]}\big)-\Omega\big(\phi^{[i]},\phi^{[i]}\big)=\big(\widetilde A^2-4\widetilde B\widetilde C\big)-\big(A^2-4BC\big),
\end{align}
which illustrates that $A^2-4BC$ remains invariant under any diffeomorphism $\exp(\xi)^*$. This provides a rigorous proof that $A^2-4BC$ is constant within an equivalence class. Now, let us elucidate the physical significance of the inner product $\Omega$ or Killing form $\mathfrak K$. For any $\phi\in\mathsf{Sol}(\phi)$, Denote $\hat\phi=\sigma^{-1}(\phi)\in\mathfrak{sl}(2,\mathbb R)$, and subsequently, 
\begin{align*}
	\hat\phi|\vartheta_A\rangle=\sigma[\hat\phi, \xi_a],\quad \hat\phi|\vartheta_B\rangle=\sigma[\hat\phi, \xi_b],\quad \hat\phi|\vartheta_C\rangle=\sigma[\hat\phi,\xi_c].
\end{align*}
In this representation, we can express $-\Omega(\phi,\phi)=-\frac{1}{4}\mathrm{tr}\,\hat\phi^2$, which can be interpreted as a mass of the configuration~\cite{gonzalez2018,iliesiu2019}. Integrating this term over $\Sigma$ produces a boundary action, with to gether with the bulk BF action recovers the Schwarzian action \cite{iliesiu2019}. We will observe later the connection between $\Omega(\phi,\phi)$ and the Schwarzian derivative. 

By straightforward calculations, it can be verified that for any dilaton profile the trace $\mathrm{tr}\,\hat\phi^{[i]}=0$ is always zero. Thus the  ''mass" $-\Omega(\phi,\phi)$ of a dilaton profile $\phi^{[i]}$ is proportional to the fluctuation of the corresponding operator, $\mathrm{tr}(\hat\phi^{[i]})^2-(\mathrm{tr}\,\hat\phi^{[i]})^2$,  which corresponds to the norm of the dilaton under the inner product $\Omega$. Therefore, we introduce an invariant characteristic constant, denoted by $\varDelta\phi^{[i]}$, to measure its average contribution, 
\begin{align}
	\frac{1}{4}\mathrm{tr}(\hat\phi^{[i]})^2=\Omega(\phi^{[i]},\phi^{[i]})=\varDelta\phi^{[i]}\cdot \big|\varDelta\phi^{[i]}\big|\ .
\end{align}
It is clear that $\varDelta\phi^{[i]}$ shares the same sign with  $\Omega(\phi^{[i]},\phi^{[i]})$. Furthermore, we can define the average potential energy of the dilaton profile
\begin{align}
	\varDelta\mathscr U^{[i]}(\phi^{[i]})=\mathscr U^{[i]}(\varDelta\phi^{[i]})\ .
\end{align}
We will delve into the role of the average potential energy in the next section.

Previously we have classified the non-trivial dilatons into three types, namely \eqref{phiA}, \eqref{phiB}, and \eqref{phiC}. The classification is based on the sign of $A^2-4BC$. For convenience, we introduce a unified parameter $K \geqslant 0$, defined by $K^2 = |A^2 - 4BC|$. It is then evident that, for \eqref{phiA}, we have $K = A$; for \eqref{phiB}, we have $K = 0$; and for \eqref{phiC}, we have $K = 2BL$, where $L$ denotes a certain $L^{[i]}$ associated with a given page. Therefore,
\begin{subequations}
	\label{classifydilaton}
	\begin{align}
		\label{classifydilaton+}
		\text{Type $+$:}\qquad   & \phi(t,z)= K\,\frac{t}{z},  \quad  K>0;                                               \\
		\label{classifydilaton0}
		\text{Type $0$\,:}\qquad & \phi(t,z)=\pm L\,\frac{1}{z}, \quad K=0;                                                        \\
		\label{classifydilaton-}
		\text{Type $-$:}\qquad   & \phi(t,z)=\pm  \dfrac{K}{2L} \bigg(\frac{t^2-z^2}{z}+ L^2\,\dfrac{1}{z}\bigg), \quad  K>0. 
	\end{align}
\end{subequations}
It should be noted that although \eqref{classifydilaton} employs a unified parameter $K$, this does not imply that every $K$ is identical; on the contrary, it depends on the specific dilaton under consideration. Even if two dilatons belong to the same type, different values of $K$ will render them inequivalent. The parameter $K$ serves as a physical characteristic of a given dilaton. Notice that not all transformaitons~\eqref{oneparatransmation} rotates amongs the solutions within a family, there are residue symmetries keeping the representatives~\eqref{classifydilaton} of each type unchanged. Explicitly, Type $+$ solution is invariant under $\exp(a\xi_a)$ transformations; Type $0$ is invariant under $\exp(c\xi_c)$; Type $-$ is invariant under $\exp\big(\theta\xi_b/L^{[i]}+\theta L^{[i]}\xi_c\big)$. Let us take the Type $-$ case as an example 
\begin{align}
	\label{Type-arctanti}
	\exp\big(\theta\xi_b/L^{[i]}+\theta L^{[i]}\xi_c\big)^*:\quad \arctan \bigg(\frac{t(\theta)\pm z(\theta)}{L^{[i]}}\bigg)= \arctan\bigg(\frac{t\pm z}{L^{[i]}}\bigg)+\theta.
\end{align}

Recall that all group elements $\kappa\in\mathrm{SL}(2,\mathbb R)$ can also be categorized into three classes~\cite{Deitmar2009}: hyperbolic, parabolic, and elliptic, corresponding to $|\mathrm{tr}\,\kappa|>2,\;=2,\;<2$, respectively.
The classification of dilatons shares similarities with the classification of group elements. However, the two classification schemes are not in one-to-one correspondence; hence, although they are related, they are essentially distinct.
 Appendix \ref{appx2} provides a detailed exploration of the correlations between them.

\subsubsection{Continuity conditions of the dilaton\label{sec4}}
We now turn to the junction conditions of the dilaton. Firstly the dilaton should be continuous on the interface, namely $\bar\phi=\phi^{[i]}\big|_\Sigma$,
$\forall i$. On every page $\mathscr V^{[i]}$, we can choose a Poincar\'e patch, allowing us to express the continuity conditions as
\begin{align}
	\label{nonperturdilaton}
	\bar\phi(u)=\phi^{[i]}\big|_\Sigma= A^{[i]}\frac{ t^{[i]}(u)}{z^{[i]}(u)}+ B^{[i]}\frac{t^{[i]}(u)^2-z^{[i]}(u)^2}{z^{[i]}(u)}+ C^{[i]}\frac{1}{z^{[i]}(u)}.
\end{align}
Similar to the computation of extrinsic curvature, our objective is to expand $\bar\phi(u)$ in terms of $\epsilon$ and then systematically analyze the continuity conditions and junction conditions at each order. Employing the expansion of $z^{[i]}(u)$~\eqref{epsilonO3}, we get
\begin{align}
	\label{phiuexpand}
	\notag
	\phi^{[i]}\big|_\Sigma= & \phi^{[i]}\big(t^{[i]}(u),z^{[i]}(u)\big)= \phi^{[i]}_{(-1)}(u)\cdot \frac{1}{\epsilon}+\phi^{[i]}_{(0)}(u)+\phi^{[i]}_{(1)}(u)\cdot\epsilon +\phi^{[i]}_{(2)}(u)\cdot\epsilon^2+\mathcal O(\epsilon^3)                                                                                           \\
	\notag
	=                       & \frac{1}{L^{[i]}}\bigg\{A^{[i]}\frac{t^{[i]}(u)}{t^{[i]\prime}(u)}+ B^{[i]}\frac{t^{[i]}(u)^2}{t^{[i]\prime}(u)}+C^{[i]}\frac{1}{t^{[i]\prime}(u)}\bigg\}\cdot\frac{1}{\epsilon}+L^{[i]}\bigg\{A^{[i]}\frac{t^{[i]}(u)t^{[i]\prime\prime}(u)^2}{2t^{[i]\prime}(u)^3} \\
	                        & + B^{[i]}\bigg(\frac{t^{[i]}(u)^2t^{[i]\prime\prime}(u)^2}{2t^{[i]\prime}(u)^3}-t^{[i]\prime}(u)\bigg)+C^{[i]}\frac{t^{[i]\prime\prime}(u)^2}{2t^{[i]\prime}(u)^3}\bigg\}\cdot\epsilon  +\mathcal O(\epsilon^3).
\end{align}
Notice that even powers of $\epsilon$, namely $\phi^{[i]}_{2(p)}$ terms, vanish in this expression.

\subsubsection{Near-boundary expansion of the dilaton's continuity conditions}
Due to the influence of lower-order behaviors on higher orders, for any odd $p\geqslant -1$, the continuity condition at the $p$-th order should take the following form,
\begin{align}
	\label{eachordercontinu}
	\sum_{\Bbbk=-1}^p \phi^{[i]}_{(\Bbbk)}(u) {\epsilon} ^\Bbbk=\sum_{\Bbbk=-1}^p \phi^{[j]}_{(\Bbbk)}(u) {\epsilon} ^\Bbbk+\mathcal O(\epsilon^{p+2}), \qquad \forall i,j\in\big\{0,\cdots,m\big\}.
\end{align}
Specifically, for $p=-1$,
\begin{align}
	\label{O(-1)continu}
	\phi^{[i]}_{(-1)}(u)=\phi^{[j]}_{(-1)}(u)+\mathcal O(\epsilon^2).
\end{align}
For $p=1$,  we have
\begin{align}
	\label{O(1)continu}
	\phi^{[i]}_{(-1)}(u)+\phi^{[i]}_{(1)}(u) {\epsilon} ^2=\phi^{[j]}_{(-1)}(u)+\phi^{[j]}_{(1)}(u)\epsilon^2+\mathcal O(\epsilon^4).
\end{align}
Clearly, when exclusively focusing on the continuity conditions at the leading order, there is no need to worry about the impact of coordinate transformations, as the differences are only of $\mathcal O(\epsilon^2)$. However, when extending the consideration to the continuity conditions for $p=1$ and $p>1$, the above equation fails to produce meaningful results, because the expressions that hold order by order still contain $\epsilon$. However, for $p=1$ and even higher orders, we need a stronger, more effective, yet still reasonable continuity condition that does not involve $\epsilon$. On one hand, this ensures that the continuity condition achieves the same level of precision as the junction condition; on the other hand, for the finite cutoff of AdS$_2$, higher-order contributions become non-negligible.

To eliminate the effect of $\epsilon$ in the continuity condition at each order, we must select a specific prior Poincar\'e coordinate $t^{[i]}(u)$ for each page. We assume that, in this coordinate, the $\mathcal O(\epsilon^2)$ term in \eqref{O(-1)continu} is strictly zero, thereby allowing the $\epsilon$ on both sides of \eqref{O(1)continu} to cancel out. That is,
\begin{subequations}
	\label{strongercontinu}
	\begin{align}
		\label{strongercontinu-1}
		&\phi^{[i]}_{(-1)}(u)=\phi^{[j]}_{(-1)}(u);\\
		\label{strongercontinu1}
		&\phi^{[i]}_{(1)}(u)=\phi^{[j]}_{(1)}(u)+\mathcal O(\epsilon^2).
	\end{align}
\end{subequations}
We can even impose a stronger condition: that $\mathcal O(\epsilon^2)$ vanishes in the continuity condition at each order, thus
\begin{align}
	\label{strongercontinuorder}
	\phi^{[i]}_{(p)}(u)=\phi^{[j]}_{(p)}(u), \quad \forall p\geqslant -1.
\end{align}
This assumption enables us to resolve the continuity condition at any order. Furthermore, if we apply the same assumption to the junction condition, we can jointly solve all orders of the continuity and junction conditions within a unified coordinate $t^{[i]}$. However, such a stringent assumption may not be necessary, as subsequent calculations indicate that the physical constraints at the subleading order alone are sufficient to determine the form of the dilaton. Additional solutions could even introduce inconsistencies. Given that the series expansion shares the same form across the $\mathrm{AN}$-class to which $t^{[i]}$ belongs, we are, in effect, selecting a class of coordinates, rather than a single one, for each page. Recalling from section \ref{sec3}, by using the invariant $(A^{[i]})^2-4B^{[i]}C^{[i]}$, we can categorize all dilatons into three distinct types. Each type of dilaton removes most of the redundant degrees of freedom associated with metric-preserving transformations, leaving only a single-parameter coordinate class. Within this coordinate class, the dilaton assumes a fixed form, as detailed in \eqref{classifydilaton}, where $A^{[i]}, B^{[i]}, C^{[i]}$ are mapped onto a physical parameter directly associated with $(A^{[i]})^2-4B^{[i]}C^{[i]}$. This coordinate class holds particular significance, making it a promising candidate for computing the series expansion. In this coordinate (class), when comparing the dilatons across different pages, the dilaton's sole physical parameter effectively captures its essential characteristics. At this stage, the continuity conditions \eqref{strongercontinu-1}, \eqref{strongercontinu1}, and even \eqref{strongercontinuorder} become justifiable.

It happens that, for the Type $+$ and Type $0$ dilatons, whose remaining single-parameter coordinate transformation degrees of freedom are $\exp(a\xi)^*$ and $\exp(c\xi)^*$, respectively, both belong to the $\mathrm{AN}$ subgroup. Therefore, when gluing some Type $+$ pages with some Type $0$ pages, we can perform the series expansion of the dilaton in the forms given by \eqref{classifydilaton+} and \eqref{classifydilaton0}. In this case, the value of $\phi^{[i]}_{(p)}(u)$ at each order is unique, and the remaining single-parameter coordinate transformations for both will not introduce $\mathcal O(\epsilon^2)$ corrections to $\phi^{[i]}_{(p)}(u)$. As a result, the continuity conditions \eqref{strongercontinu-1} and \eqref{strongercontinu1} behave well. Subsequent detailed calculations confirm that for the gluing of Type $+$ and Type $-$, this procedure indeed leads to the correct physical conclusions. Given the validity of the continuity condition \eqref{strongercontinuorder}, we can define the boundary values of the dilaton order by order, 
\begin{align}
	\label{expanboundarydilaton}
	\bar\phi_{(p)}(u)\coloneqq \phi^{[i]}_{(p)}(u)=\phi^{[j]}_{(p)}(u),\qquad \forall p\geqslant -1
\end{align}
where each $\mathscr V^{[i]}$ belongs to either Type $+$ or Type $0$. In this case, the series expansion of the dilaton $\phi^{[i]}$ on each page has a certain sense of uniqueness, and this uniqueness defines the expansion of the boundary dilaton $\bar\phi(u)$:
\begin{align}
	\label{expandcontinu}
	\bar\phi(u)=\bar\phi_{(-1)}(u)\cdot\frac{1}{\epsilon}+\bar\phi_{(1)}(u)\cdot\epsilon+\mathcal O(\epsilon^3).
\end{align}
In other literature, $\bar\phi_{(-1)}(u)$ is commonly denoted by the symbol $\phi_r(u)$~\cite{sarosi2017,Luca2020,Maldacena:2016upp}.

However, if we wish to further glue a Type $-$ page with pages of the other two types, the situation changes fundamentally. Let the Type $-$ page be denoted as $\mathscr V^{[k]}$. We can choose a one-parameter coordinate family, denoted as $\tilde t^{[k]}_\theta$, such that the Type $-$ dilaton takes the form given in \eqref{classifydilaton-}, where the only remaining parameter $B^{[k]}$ is a physical parameter. For any such coordinate $t^{[k]}$, we associate it with $\theta=0$, so that $\tilde t^{[k]}_\theta=\exp\big(\theta\xi_b/L^{[k]}+\theta L^{[k]}\xi_c\big)^*t^{[k]}$. However, the one-parameter metric-preserving  transformation $\exp\big(\theta\xi_b/L^{[k]}+\theta L^{[k]}\xi_c\big)^*$ does not belong to the $\mathrm{AN}$-subgroup, meaning that this transformation will introduce $\mathcal O(\epsilon^2)$ corrections to $t^{[k]}$. As a result, each $\tilde t^{[k]}_\theta$ leads to a completely different series expansion for the dilaton. For the gluing of Type $-$ pages with other types, or even the gluing of all pages as Type $-$, this non-uniqueness makes it difficult to apply the continuity condition \eqref{strongercontinuorder}. Nevertheless, we cannot rule out the possibility that a page contains a Type $-$ dilaton, so we are forced to completely break the coordinate transformation symmetry of the Type $-$ dilaton. This means we assume the existence of a unique special $\theta^{[k]}$, and expand $\phi^{[k]}$ in the unique coordinate system $\tilde t^{[k]}_{\theta^{[k]}}$. We then require the expansion terms to satisfy the continuity conditions \eqref{strongercontinu-1} and \eqref{strongercontinu1}, and even \eqref{strongercontinuorder}. In this case, \eqref{expandcontinu} can hold, i.e., $\bar\phi_{(p)}(u)=\phi^{[k]}_{(p)}(u)$ with $\mathcal O(\epsilon^2)$ strictly vanishing. The parameter $\theta^{[k]}$ can be treated as an undetermined parameter, whose value can be constrained through actual calculations. But subsequent results indicate that such a $\theta^{[k]}$ does not exist, revealing that for Type $-$ dilatons, only the leading-order continuity condition has physical significance. Alternatively, Type $-$ dilatons can never exhibit consistent continuity with other types at higher orders. We will gradually demonstrate this in the following sections.

\subsubsection{The junction condition of the dilaton}
Now we examine the dilaton's junction condition. Combining~\eqref{JTjunctionPhi} and \eqref{AdS2tension}, we can express the junction condition as
\begin{align}
	\label{AdS2dilatonjunction}
	0=\chi\bar\phi+\sum_{i=0}^m \mathfrak L_{n^{[i]}}\phi^{[i]}=\sum_{i=0}^m \mathfrak L_{n^{[i]}}\phi^{[i]}+\frac{1}{L^{[i]}}\phi^{[i]}\big|_\Sigma.
\end{align}
On each page, we choose a Poincar\'e coordinate system $(t^{(i)}, z^{(i)})$, the normal vector takes the form~\eqref{normalvecforAdS2} and the general expression \eqref{nonperturdilaton} for the dilaton, we obtain
\begin{align}
	\notag
	\mathfrak L_{n^{[i]}}\phi^{[i]}=\epsilon\bigg\{ &  A^{[i]}\biggl(\frac{z^{[i]\prime}(u)}{z^{[i]}(u)}-\frac{t^{[i]\prime}(u)t^{[i]}(u)}{z^{[i]}(u)^2}\biggr)- C^{[i]}\frac{t^{[i]\prime}(u)}{z^{[i]}(u)^2} \\
     & + B^{[i]}\bigg(\frac{2t^{[i]}(u)z^{[i]\prime}(u)}{z^{[i]}(u)}-\frac{t^{[i]\prime}(u)t^{[i]}(u)^2}{z^{[i]}(u)^2}-t^{[i]\prime}(u)\bigg)\bigg\}.
\end{align}
Employing~\eqref{epsilonO3} to expand $\mathfrak L_{n^{[i]}}\phi^{[i]}$, we get
\begin{align}
	\notag
	 & \mathfrak L_{n^{[i]}}\phi^{[i]}=\partial_\ell\phi^{[i]}_{(-1)}(u)\cdot\frac{1}{ {\epsilon} }+\partial_\ell\phi^{[i]}_{(1)}(u)\cdot {\epsilon} +\mathcal O( {\epsilon} ^2)                                                                                                                                                                 \\
	\notag
	 & =-\frac{1}{(L^{[i]})^2}\bigg\{ A^{[i]}\frac{t^{[i]}(u)}{t^{[i]\prime}(u)}+ B^{[i]}\frac{t^{[i]}(u)^2}{t^{[i]\prime}(u)}+ C^{[i]}\frac{1}{t^{[i]\prime}(u)}\bigg\}\cdot\frac{1}{ \epsilon }+\bigg\{ A^{[i]}\bigg(\frac{t^{[i]\prime\prime}(u)}{t^{[i]\prime}(u)}-\frac{t^{[i]}(u)t^{[i]\prime\prime}(u)^2}{t^{[i]\prime}(u)^3}\bigg) \\
	 & \quad+ B^{[i]}\bigg(\frac{2t^{[i]}(u)t^{[i]\prime\prime}(u)}{t^{[i]\prime}(u)}-\frac{t^{[i]}(u)^2t^{[i]\prime\prime}(u)^2}{t^{[i]\prime}(u)^3}-t^{[i]\prime}(u)\bigg)- C^{[i]}\frac{t^{[i]\prime\prime}(u)^2}{t^{[i]\prime}(u)^3}\bigg\}\cdot {\epsilon} +\mathcal O(\epsilon^3)\ .
\end{align}
Together with the expansion of $\phi^{[i]}\big|_\Sigma$~\eqref{phiuexpand}, at the leading order, we can obtain the following coordinate-independent equation:
\begin{align}
	\label{leadingjunctioni}
	\partial_\ell\phi^{[i]}_{(-1)}(u)+\frac{1}{L^{[i]}}\phi^{[i]}_{(-1)}(u)=0.
\end{align}
Summing the above equation over $[i]$, we can observe that the junction condition \eqref{AdS2dilatonjunction} automatically holds at the leading order. 
From eq.\eqref{partialF}, we have known that the leading-order junction conditions must be satisfied. Fortunately, eq.\eqref{leadingjunctioni} shows that the $\mathcal O(\epsilon^2)$ term in \eqref{partialF} is strictly zero, so under coordinate transformations, the subleading order $p=1$ remains unaffected by the leading order $p=-1$. Therefore, the subleading-order junction condition also holds physical significance. 
However, for the $p=1$ order, $\partial_\ell\phi^{[i]}_{(1)}(u)+\frac{1}{L^{[i]}}\phi^{[i]}_{(1)}(u)$ clearly depends on the choice of $t^{[i]}(u)$. Therefore, when certain pages undergo coordinate transformations, the corresponding junction condition will introduce corrections of $\mathcal O(\epsilon^2)$,
\begin{align}
	\label{p=1junction}
	\sum_{i=0}^m\partial_\ell\phi^{[i]}_{(1)}(u)+\frac{1}{L^{[i]}}\phi^{[i]}_{(1)}(u)=\mathcal O(\epsilon^2).
\end{align}
Then, for the junction condition at $p=3$ order, there will be $\mathcal O(1)$ corrections. In principle, similar to the approach for continuity conditions, we can utilize eq.\eqref{classifydilaton} to eliminate the coordinate transformation degrees of freedom for the dilaton. At this point, we can assume that the junction condition holds order by order as follows:
\begin{align}
	\label{juncconatporder}
	\sum_{i=0}^m\partial_\ell\phi^{[i]}_{(p)}(u)+\frac{1}{L^{[i]}}\phi^{[i]}_{(p)}(u)=0,\qquad \forall p\geqslant -1,
\end{align}
such that we can continue to solve for higher-order constraints. 
Recalling eq.\eqref{SchtO2}, we observe that, under coordinate transformations, for orders $p>2$, the junction condition for the extrinsic curvature will also introduce $\mathcal O(1)$ corrections. Therefore, similar operations can be carried out. However, the assumption above is too strong and may lead to incompatible conclusions. For simplicity and naturalness, in this paper, we only consider cases up to subleading order for both junction and continuity conditions.

In the following discussion, we shall no longer consider the general form of the dilaton that simultaneously involves $A^{[i]}, B^{[i]}, C^{[i]}$. Instead, we will employ the representation given in \eqref{classifydilaton}, namely the dilaton with redundant degrees of freedom eliminated. This greatly simplifies the otherwise complicated solving procedure. Since the gluing of dilatons of different types is involved, the unified parameter $K$ in \eqref{classifydilaton} may lead to confusion. Therefore, for type $+$ we replace $K$ by the notation $A^{[i]}$, while for type $-$ we substitute $\frac{K}{2L}$ with $B^{[i]}$. Now, we can impose constraints on the coordinate $t^{[i]}$ such that each type of dilaton retains the form given in \eqref{classifydilaton}. 

We then explicitly solve the constraints imposed by the leading-order continuity condition \eqref{strongercontinu-1} or \eqref{expanboundarydilaton} with $p=-1$,
\begin{align}
	\bar\phi_{(-1)}(u)=\phi^{[i]}_{(-1)}(u)=\begin{cases}
		                                        \vspace{12pt}
		                                        \dfrac{ A^{[i]}}{L^{[i]}}\dfrac{t^{[i]}(u)}{t^{[i]\prime}(u)},  \quad  A^{[i]}>0, \quad                  & \text{Type $+$};   \\
		                                        \vspace{12pt}
		                                        \pm \dfrac{1}{t^{[i]\prime}(u)},  \quad                                                                & \text{Type $0$\,}; \\
		                                        \pm\dfrac{ B^{[i]}}{L^{[i]}}\cdot\dfrac{t^{[i]}(u)^2+(L^{[i]})^2}{t^{[i]\prime}(u)},  \quad  B^{[i]}>0 , & \text{Type $-$}.    \\
	                                        \end{cases}
\end{align}
First for Type $+$, solving for $t^{[i]}(u)$ results in
\begin{align}
	\label{type+ti}
	\text{Type $+$}:\qquad t^{[i]}_{\pm}(u)=\pm L^{[i]}\cdot\exp\bigg(\frac{ A^{[i]}}{L^{[i]}}\int\frac{1}{\bar\phi_{(-1)}(u)}\,\mathrm du\bigg).
\end{align}
The choice of $\pm$ must ensure that $t^{[i]\prime}(u)>0$ because both the time coordinate $u$ and $t^{[i]}$ should point towards the future. Hence, when $\bar\phi_{(-1)}(u)>0$, we select $t^{[i]}_+(u)$; and conversely, we opt for $t^{[i]}_-(u)$. However, in practice, it can be ascertained through subsequent calculations that the sign of $t^{[i]}_{\pm}$ does not impact the main conclusions. Because the junction conditions remain invariant under the discrete transformation $t^{[i]}\to -t^{[i]}$. The expression of $t^{[i]}(u)$ is invariant under $\exp(a\xi_a)^*$, where $a$ precisely corresponds to the integration constant arising from the indefinite integral $\int\frac{1}{\bar\phi_{(-1)}(u)}\,\mathrm du$.

In the case of Type $0$, solving for $ t^{[i]}(u)$,
\begin{align}
	\label{type0ti}
	\text{Type $0$\,}:	\qquad t^{[i]}_{\pm}(u)=\pm \int\frac{1}{\bar\phi_{(-1)}(u)}\,\mathrm du.
\end{align}
Here, when $\bar\phi_{(-1)}(u)>0$, we choose $t^{[i]}_+(u)$; and conversely, choose $t^{[i]}_-(u)$. The parameter $c$ for the time translation directly corresponds to the integration constant produced by the indefinite integral $\int\frac{1}{\bar\phi_{(-1)}(u)}\,\mathrm du$.

Lastly, for Type $-$, solving for $t^{[i]}(u)$,
\begin{align}
	\label{type-ti}
	\text{Type $-$}:\qquad t^{[i]}_{\pm}(u)=\pm L^{[i]}\tan\bigg(B^{[i]}\int \frac{1}{\bar\phi_{(-1)}(u)}\,\mathrm du+\theta^{[i]}\bigg).
\end{align}
When $\bar\phi_{(-1)}(u)>0$, we opt for $t^{[i]}_+(u)$; and when $\bar\phi_{(-1)}(u)<0$, choose $t^{[i]}_-(u)$. In the above expression, we introduce the parameter $\theta^{[i]}$ to be solved, aiming to offset the integration constant generated by $\int\frac{1}{\bar\phi_{(-1)}(u)}\,\mathrm du$. Following the assumption introduced in our previous discussion, the dilaton of Type $-$ has completely lost its coordinate transformation degrees of freedom. According to \eqref{Type-arctanti}, the residual one-parameter transformation $\exp\bigl(\theta\xi_b/L^{[i]}+\theta L^{[i]}\xi_c\bigr)^*$ acting on $t^{[i]}(u)$ results in:
\begin{align}
	\label{tantheta}
	\exp\bigl(\theta\xi_b/L^{[i]}+\theta L^{[i]}\xi_c\bigr)^*t^{[i]}= L^{[i]}\tan\bigg(\arctan \frac{t^{[i]}}{L^{[i]}}+\theta\bigg)+\mathcal O(\epsilon^2).
\end{align}
We observe that $\exp\bigl(\theta\xi_b/L^{[i]}+\theta L^{[i]}\xi_c\bigr)^*$ precisely induces a phase shift of $\theta$ for $t^{[i]}(u)$. Therefore, introducing a fixed phase $\theta^{[i]}$ is a manifestation of the complete violation of coordinate symmetry.

\subsection{Schwarzian derivative and $\mathfrak{sl}(2,\mathbb{R})$}
We temporarily interrupt the discussion on the junction condition. Utilizing the expression for $ t^{[i]}(u)$, we can examine the Schwarzian theory at the boundary. Starting from the JT gravity action \eqref{JTaction} and substituting the AdS$_2$ metric, only the boundary terms remain. Then substituting  the expansions of $\bar\phi(u)$ and $K^{[i]}$ with respect to $\epsilon$,
\begin{align}
	16\pi G_{\mathrm N}\,I_{\mathscr V^{[i]}}=-2\oint_\Sigma\bar\phi(u) K^{[i]}\frac{\mathrm du}{\epsilon}=\frac{2}{\epsilon^2L^{[i]}}\oint_\Sigma\bar\phi_{(-1)}(u)\,\mathrm du+16\pi G_{\mathrm N}\, I^{[i]}_{\text{Sch}}+\mathcal O(\epsilon^2),
\end{align}
where $I^{[i]}_{\text{Sch}}$ represents the Schwarzian action of page $\mathscr V^{[i]}$ on the boundary,
\begin{align}
	\label{Schaction}
	I^{[i]}_{\text{Sch}}=-\frac{L^{[i]}}{8\pi G_{\mathrm N}}\oint_\Sigma\bigg\{\bar\phi_{(-1)}(u)\mathsf{Sch}[t^{[i]}](u)-\frac{1}{(L^{[i]})^2}\bar\phi_{(1)}(u)\bigg\}\,\mathrm du.
\end{align}
It can be observed that, compared to the familiar form~\cite{iliesiu2019,mertens2023,sarosi2017,engelsoy2016inves,stanford2017fermionic,Maldacena:2016upp} of the Schwarzian action, there is an additional correction involving $\bar\phi_{(1)}(u)$. This is due to our consideration of higher-order terms in the expansion of the dilaton. Fortunately, in action \eqref{Schaction}, $\bar\phi_{(1)}(u)$ is not coupled with the dynamical variable $t^{[i]}(u)$, so the action \eqref{Schaction} must yield the same dynamics as the classical form of the Schwarzian action. Upon varying \eqref{Schaction}, the following equation \cite{Maldacena:2016upp} is obtained,
\begin{align}
	\bigg\{\frac{1}{t^{[i]\prime}}\bigg(\frac{\bigl(t^{[i]\prime}\bar\phi_{(-1)}\bigr)^\prime}{t^{[i]\prime}}\bigg)^\prime\bigg\}^\prime=0.
\end{align}
By substituting the expressions for $ t^{[i]}(u)$ of Type $+$, Type $0$, and Type $-$ from eqs.\eqref{type+ti}, \eqref{type0ti}, and \eqref{type-ti} into the above equation, we derive only three identities. Consequently, all the $ t^{[i]}(u)$ under scrutiny must indeed be solutions to the boundary Schwarzian theory.

Now we calculate the Schwarzian derivative $\mathsf{Sch}\big[t^{[i]}_{\pm}\big]$ separately for the three different types. Since the Schwarzian derivative in different coordinates only differs at higher orders, we can anticipate that $\mathsf{Sch}\big[t^{[i]}_{\pm}\big]$ does not depend on $\int\frac{1}{\bar\phi_{(-1)}(u)}\,\mathrm du$. In particular, because the Schwarzian derivative remains invariant under the discrete transformation $t\to -t$, the sign of $t^{[i]}_{\pm}$ does not affect its value. In fact, we have
\begin{align}
	\mathsf{Sch}\big[t^{[i]}_{\pm}\big](u)=\frac{\bar\phi_{(-1)}^{\,\prime}(u)^2-2\bar\phi_{(-1)}(u)\bar\phi_{(-1)}^{\,\prime\prime}(u)+ D^{[i]}}{2\bar\phi_{(-1)}(u)^2},
\end{align}
where the constants $ D^{[i]}$ for dilatons of different types are defined as
\begin{align}
	\label{Di}
	 D^{[i]}=\begin{cases}
		        \vspace{9pt}
		        -\bigg(\dfrac{ A^{[i]}}{L^{[i]}}\bigg)^2, & \text{Type $+$};   \\
		        \vspace{9pt}
		        0,                                       & \text{Type $0$\,}; \\
		        4( B^{[i]})^2,                            & \text{Type $-$}.
	        \end{cases}
\end{align}
All $\mathsf{Sch}[t^{[i]}]$ shares a common portion that is solely determined by $\bar\phi_{(-1)}(u)$ and its derivative. The constants $ D^{[i]}$ depend on the geometric parameters and the type of the page, reflecting the influence of the dilaton outside $\Sigma$. We explain the physical significance of $D^{[i]}$. 

In section \ref{sec3}, we introduced the inner product $\Omega$ of the dilaton space, and using this inner product,  defined an mean constant $\Delta\phi^{[i]}$ of the dilaton on $\mathscr V^{[i]}$, which satisfies $\Omega(\phi^{[i]},\phi^{[i]})=\Delta\phi^{[i]}\cdot \left|\Delta\phi^{[i]}\right|$. Therefore, $ D^{[i]}=-2\Omega(\phi^{[i]},\phi^{[i]})/(L^{[i]})^2=\mathscr U^{[i]}\big(\Delta\phi^{[i]}\big)\cdot\left|\Delta\phi^{[i]}\right|$. In other words, $ D^{[i]}$ is proportional to the average potential energy of the dilaton. When $ D^{[i]}<0$, which means the average potential energy is negative, it represents an attractive action of the dilaton $\phi^{[i]}$. When $ D^{[i]}=0$, the average potential energy vanishes, and $\phi^{[i]}$ is neutral. When $ D^{[i]}>0$, positive potential energy reflects the repulsive effect of the dilaton. We will examine these properties in more detail later. Now, we define a constant $E$ that represents the overall propertie of the booklet $\mathscr P$,
\begin{align}
	\label{E}
	E=\frac{\sum_{i=0}^m  D^{[i]}L^{[i]}}{\sum_{i=0}^m L^{[i]}}.
\end{align}
In other words, $E$ is the weighted arithmetic average of all $ D^{[i]}$ with respect to the AdS radius, representing the total effects of attraction and repulsion of the dilatons in all pages. Its magnitude reflects the proportions of various types of pages within the booklet $\mathscr P$. Specifically, if all pages are of Type $+$, then $E$ is necessarily less than $0$; if all are of Type $0$, then $E=0$; and lastly, if they are all of Type $-$, then $E>0$.

The junction condition for extrinsic curvature yields Eq.\eqref{sumSchLi}, which states that the sum of all $L^{[i]}\mathsf{Sch}[t^{[i]}]$ must vanish, then
\begin{align}
	\bar\phi_{(-1)}^{\,\prime}(u)^2-2\bar\phi_{(-1)}(u)\bar\phi_{(-1)}^{\,\prime\prime}(u)+E=0,
\end{align}
which allows us to determine the form of $\bar\phi_{(-1)}(u)$,
\begin{align}
	\label{phi(-1)u}
	\bar\phi_{(-1)}(u)=\frac{E+\alpha^2}{4\beta}u^2+\alpha u+\beta,
\end{align}
where $\alpha$ and $\beta$ are integration constants. Only when the $u^2$ term disappears can $\beta=0$ exist. Later on, we will independently derive the form of $\bar\phi_{(-1)}(u)$ from the continuity condition and the junction condition for the dilaton, and compare the results with the above equation to verify the compatibility. This also serves as a verification of the assumption underlying the subleading-order continuity condition \eqref{strongercontinu1}.  

As a side remark, the Schwarzian derivative $\mathsf{Sch}[t^{[i]}]$ is known to be invariant under $\mathrm{SL}(2,\mathbb R)$, therefore the junction condition~\eqref{sumSchLi} is also invariant. 

 $\exp(a\xi_a)^*$ and $\exp(c\xi_c)^*$. For the remaining $\exp\big(\theta(\xi_b/L^{[i]}+L^{[i]}\xi_c)\big)^*$, utilizing Eq.\eqref{Type-arctanti}, direct calculations reveal that,
\begin{align}
	\label{SchtO2}
	\mathsf{Sch}[\exp\big(\theta\xi_b/L^{[i]}+\theta L^{[i]}\xi_c\big)^*t^{[i]}]-\mathsf{Sch}[t^{[i]}]=\mathcal O(\epsilon^2).
\end{align}
It is evident that the metric-preserving transformation does not lead to corrections at the same order in \eqref{sumSchLi}, making it a meaningful constraint. However, for the junction condition of the extrinsic curvature, the expansion term at $\mathcal O(\epsilon^4)$ does depend on the choice of $t^{[i]}(u)$.

\subsection{Solving the subleading-order conditions}\label{subsec4E}
Now, let us examine the continuity condition at the $p=1$ order. For any page $\mathscr V^{[i]}$ that is either Type $+$ or Type $0$, eq.\eqref{strongercontinu1} implies
\begin{align}
	\label{phi1(u)allType}
	\bar\phi_{(1)}(u)=\phi^{[i]}_{(1)}(u)=\begin{cases}
		                                      \vspace{1.5ex}
		                                      L^{[i]} A^{[i]}\dfrac{t^{[i]}_{\pm}(u)t^{[i]\prime\prime}_{\pm}(u)^2}{2t^{[i]\prime}_{\pm}(u)^3}=\dfrac{\big( A^{[i]}-L^{[i]}\bar\phi_{(-1)}^{\,\prime}(u)\big)^2}{2\bar\phi_{(-1)}(u)}, & \text{Type $+$};  \\
		                                      \pm(L^{[i]})^2\dfrac{t^{[i]\prime\prime}_{\pm}(u)^2}{2t^{[i]\prime}_{\pm}(u)^3}=(L^{[i]})^2\dfrac{\bar\phi_{(-1)}^{\,\prime}(u)^2}{2\bar\phi_{(-1)}(u)},                               & \text{Type $0$}. \\
	                                      \end{cases}
\end{align}
From the above expressions, we see that for these two types, $\bar\phi_{(1)}(u)$ locally depends on $\bar\phi_{(-1)}(u)$ and $\bar\phi_{(-1)}^{\,\prime}(u)$, while being independent of the integral $\int \frac{\mathrm du}{\bar\phi_{(-1)}(u)}$. This precisely reflects the invariance of $\bar\phi_{(1)}(u)$ under the residual transformations. But for Type $-$, the continuity condition at the $p=1$ order yields
\begin{align}
	\label{phi1(u)Type-}
	\notag
	 & \text{Type $-$}:\quad \bar\phi_{(1)}(u)=\phi^{[i]}_{(1)}(u)=\pm L^{[i]} B^{[i]}\bigg(\big(t^{[i]}_{\pm}(u)^2+(L^{[i]})^2\big)\cdot\dfrac{t^{[i]\prime\prime}_{\pm}(u)^2}{2t^{[i]\prime}_{\pm}(u)^3}-t^{[i]\prime}_{\pm}(u)\bigg)                                               \\
	 & =\frac{( B^{[i]} L^{[i]})^2}{\bar\phi_{(-1)}(u)}\bigg\{\bigg[\tan\bigg( B^{[i]}\int \frac{\mathrm du}{\bar\phi_{(-1)}(u)}+\theta^{[i]}\bigg)-\frac{\bar\phi_{(-1)}^{\,\prime}(u)}{ B^{[i]}}\bigg]^2-\frac{1}{2}\bigg(\frac{\bar\phi_{(-1)}^{\,\prime}(u)}{ B^{[i]}}\bigg)^2-1\bigg\}.
\end{align}
Obviously, $\bar\phi_{(1)}(u)$ depends on the integral $\int \frac{\mathrm du}{\bar\phi_{(-1)}(u)}$. We recognize the necessity of introducing $\theta^{[i]}$ to offset the integration constant, which is a manifestation of fully fixing the coordinate system. All the results of three types directly demonstrate that the sign of $t^{[i]}_{\pm}$ does not affect the discussion of continuity conditions. Subsequent calculations will also show that the sign does not affect the junction condition, so we no longer label $\pm$.

On the other hand, Eq.\eqref{p=1junction} can provide the junction condition of dilatons at the $p=1$ order. We calculate the explicit expression for $\partial_\ell\phi^{[i]}_{(1)}(u)$ involved in it,
\begin{align}
	\label{pdellphii(u)forallType}
	\partial_\ell\phi^{[i]}_{(1)}(u)=\begin{cases}
		                             A^{[i]}\dfrac{t^{[i]\prime}(u)^2t^{[i]\prime\prime}(u)-t^{[i]}(u)t^{[i]\prime\prime}(u)^2}{t^{[i]\prime}(u)^3}=\dfrac{\bar\phi_{(-1)}^{\,\prime}(u)\big( A^{[i]}-L^{[i]}\bar\phi_{(-1)}^{\,\prime}(u)\big)}{\bar\phi_{(-1)}(u)}, & \text{Type $+$};   \\
		                            \vspace{1.5ex}
		                            -L^{[i]}\dfrac{t^{[i]\prime\prime}(u)^2}{t^{[i]\prime}(u)^3}=-L^{[i]}\dfrac{\bar\phi_{(-1)}^{\,\prime}(u)^2}{\bar\phi_{(-1)}(u)},                                                                                               & \text{Type $0$\,}; \\
		                            \vspace{1.5ex}
		                             B^{[i]}\bigg(\dfrac{2t^{[i]}(u)t^{[i]\prime\prime}(u)}{t^{[i]\prime}(u)}-t^{[i]\prime}(u)-\big(t^{[i]}(u)^2+(L^{[i]})^2\big)\dfrac{t^{[i]\prime\prime}(u)^2}{t^{[i]\prime}(u)^3}\bigg),                                         & \text{Type $-$},
	                            \end{cases}
\end{align}
where, for Type $-$, it can be further expressed as
\begin{align}
	\label{pdellphii(u)forType-}
	\partial_\ell\phi^{[i]}_{(1)}(u)=-\frac{L^{[i]}( B^{[i]})^2}{\bar\phi_{(-1)}(u)}\bigg\{ \bigg[\tan\bigg( B^{[i]}\int\frac{\mathrm du}{\bar\phi_{(-1)}(u)}+\theta^{[i]}\bigg)-\frac{\bar\phi_{(-1)}^{\,\prime}(u)}{ B^{[i]}}\bigg]^2+1\bigg\}.
\end{align}

Gluing together pages of different types with different numbers can lead to various possibilities, each with a distinct manifestation of continuity conditions. In other words, the constraints on $\bar\phi_{(-1)}(u)$ differ in each case. Additionally, we need to demonstrate that these constraints  are compatible with Eq.\eqref{phi(-1)u}. Before delving into a general discussion, we first examine some simple cases to gather as much information as possible. We first assume that all pages have the same type. This does not imply a requirement of $\mathbb Z_{m+1}$ symmetry because the AdS radius $L^{[i]}$ and the parameters of the dilaton, $ A^{[i]}$ or $B^{[i]}$, can vary for each page.

\subsubsection{All pages taking Type $+$}
Now we assume that the dilatons in all pages are of Type $+$, but the parameters $ A^{[i]}$ of the dilaton or the AdS radius $L^{[i]}$ cannot be entirely the same. We select two pages, $\mathscr V^{[j]}$ and $\mathscr V^{[k]}$, and require that $ L^{[j]}\not= L^{[k]}$ or $ A^{[j]}\not= A^{[k]}$ with at least one of these conditions being met. The continuity condition implies that
\begin{align}
	\label{alltype+phi1}
	\bar\phi_{(1)}(u)=\frac{\big( A^{[j]}- L^{[j]}\bar\phi_{(-1)}^{\,\prime}(u)\big)^2}{2\bar\phi_{(-1)}(u)}=\frac{\big( A^{[k]}- L^{[k]}\bar\phi_{(-1)}^{\,\prime}(u)\big)^2}{2\bar\phi_{(-1)}(u)}
\end{align}
Therefore, we must have $\bar\phi_{(-1)}^{\,\prime}(u)=\frac{A^{[j]}+ A^{[k]}}{ L^{[j]}+ L^{[k]}}$ or $\frac{ A^{[j]}- A^{[k]}}{ L^{[j]}- L^{[k]}}$, which will be a constant, denoted as $\alpha$. We can assert that $\alpha\not =0$, as otherwise, the above equation implies $ A^{[i]}=A$ for all $i$, and immediately
\begin{align}
	\bar\phi_{(-1)}(u)=\beta, \quad \bar\phi_{(1)}(u)=\frac{A^2}{2\beta}, \quad \partial_\ell\phi^{[i]}_{(1)}(u)\propto \bar\phi_{(-1)}^{\,\prime}(u)=0,
\end{align}
where $\beta$ is a non-zero constant. The subleading-order junction condition will yield $A^2=0$, which contradicts the premise of $A>0$.

Therefore, we assume $\bar\phi_{(-1)}(u)=\alpha(u+\gamma)$. In fact, since $u$ has a translational degree of freedom, we can further assume $\gamma=0$. For each $[i]$, $\big( A^{[i]}-\alpha L^{[i]}\big)^2$ is constant, and we can write $ A^{[i]}-\alpha L^{[i]}=\pm\alpha\zeta$. Then, along with eqs.\eqref{alltype+phi1} and \eqref{type+ti}, we have
\begin{align}
	\label{alltype+ti(u)}
	\bar\phi_{(-1)}(u)=\alpha(u+\gamma),\quad \bar\phi_{(1)}(u)=\frac{\alpha\zeta^2}{2(u+\gamma)},\quad t^{[i]}(u)=L^{[i]}\bigg(\frac{u+\gamma}{L^{[i]}}\bigg)^{1\pm \zeta/L^{[i]}}.
\end{align}
Here, we have already eliminated the residual one-parameter transformation degrees of freedom for $t^{[i]}(u)$. If for some pages, $1\pm \zeta/L^{[i]}<0$, then the above expression for $t^{[i]}(u)$ should have an additional negative sign. Next, we solve the constraints imposed by the dilaton's $p=1$ order junction condition. We assume that there are $n_1$ pages satisfying $ A^{[i]}-\alpha L^{[i]}=\alpha\zeta$, and thus, there are $n_2=m+1-n_1$ pages satisfying $ A^{[i]}-\alpha L^{[i]}=-\alpha\zeta$. If $\zeta=0$, $n_1$ can take any integer between $0$ and $m+1$. The junction condition gives
\begin{align}
	-\frac{\chi\alpha\zeta^2}{2(u+\gamma)}=-\chi\bar\phi_{(1)}(u)=\sum_{i=0}^m\partial_\ell \phi^{[i]}_{(1)}(u) =\frac{\big(n_1-n_2\big)\alpha\zeta}{u+\gamma},
\end{align}
namely,
\begin{align}
	\label{alltype+junction}
	\chi\zeta^2+2\big(n_1-n_2\big)\zeta=0.
\end{align}
So the junction condition restricts that $\zeta$ can only be either $0$ or $-\frac{\chi}{2(n_1-n_2)}$.

We also need to verify the compatibility of the above results with Eq.\eqref{phi(-1)u}. We calculate the constant $E$ in this scenario,
\begin{align}
	\notag
	E & =-\alpha^2\bigg(\sum_{i=0}^m  L^{[i]}\bigg)^{-1}\bigg(\sum_j\frac{\big( L^{[j]}+\zeta\big)^2}{ L^{[j]}}+\sum_k\frac{\big( L^{[k]}-\zeta\big)^2}{ L^{[k]}}\bigg) \\
	  & =-\alpha^2-\alpha^2\bigg(\sum_{i=0}^m  L^{[i]}\bigg)^{-1}\bigg(2\zeta\big(n_1-n_2\big)+\zeta^2\chi\bigg),
\end{align}
where $j$ sums over all indices that satisfy $ A^{[j]} /\alpha= L^{[j]}+\zeta$, while $k$ sums over all indices that satisfy $ A^{[k]} /\alpha= L^{[k]}-\zeta$. In conjunction with Eq.\eqref{alltype+junction}, we can obtain $E=-\alpha^2$. Comparing with Eq.\eqref{phi(-1)u}, since $\bar\phi_{(-1)}=\alpha(u+\gamma)$, we once again get $E+\alpha^2=0$, which shows that all of the junction and continuity conditions are compatible.

Regarding the dilaton, we would like to provide some comments. The internal coordinate $u$ of $\Sigma$ possesses a reparameterization symmetry, specifically, a translational symmetry. This leads us to observe that in the case of gluing $m+1$ Type $+$ pages together, there exists only one free parameter, namely $\alpha$, which represents the characteristic scale of the dilaton. By utilizing the solution for the boundary dilaton $\bar\phi(u)$, we can examine the gravitational part of the action for each page,
\begin{align}
	16\pi G_{\mathrm N}\,I_{\mathscr V^{[i]}}=\frac{2\alpha}{\epsilon^2L^{[i]}}\oint_\Sigma u\,\mathrm du+\frac{2\alpha\zeta(\zeta\pm L^{[i]})}{L^{[i]}}\oint_\Sigma \frac{1}{u}\,\mathrm du+\mathcal O(\epsilon^2),
\end{align}
where the second integral represents the Schwarzian action. It is evident that the action's value is predominantly influenced by $\alpha$. Note that $\zeta$ is contingent upon the tension $\chi$.

In particular, let us examine the situation in which $\mathscr P$ exhibits a $\mathbb Z_{m+1}$ symmetry. This means that for any index $[i]$, we have $L^{[i]}=L$ and $ A^{[i]}=A$. Consequently, we straightforwardly determine that $\chi=\frac{m+1}{L}$. The continuity condition is automatically met,
\begin{align}
	\bar\phi_{(1)}(u)=\frac{\big(A -L\,\bar\phi_{(-1)}^{\,\prime}(u)\big)^2}{2\bar\phi_{(-1)}(u)},
\end{align}
so no additional constraints are imposed. Let us proceed with the direct computation of the junction condition at $p=1$ order,
\begin{align}
	-\frac{m+1}{L}\frac{\big(A -L\,\bar\phi_{(-1)}^{\,\prime}(u)\big)^2}{2\bar\phi_{(-1)}(u)}=(m+1)\frac{\bar\phi_{(-1)}^{\,\prime}(u)\big(A -L\,\bar\phi_{(-1)}^{\,\prime}(u)\big)}{\bar\phi_{(-1)}(u)},
\end{align}
then
\begin{align}
	\bar\phi_{(-1)}^{\,\prime}(u)=\pm\frac{A}{L},\quad  \bar\phi_{(-1)}(u)=\pm\frac{A}{L}(u+\gamma),\quad t(u)=u+\gamma, \quad\text{or}\quad  t(u)=-\frac{L^2}{u+\gamma}.
\end{align}
On the other hand, $E=-\frac{A^2}{L^2}$, which is consistent with the results mentioned above.

\subsubsection{All pages taking Type $0$}
Now, we assume that all the pages have a Type $0$ dilaton, but their AdS radii are not necessarily equal. Thus, for specific indices $[j]$ and $[k]$, let $ L^{[j]}\not= L^{[k]}$. In this scenario, the continuity condition
\begin{align}
	\bar\phi_{(1)}(u)=( L^{[j]})^2\frac{\bar\phi_{(-1)}^{\,\prime}(u)^2}{2\bar\phi_{(-1)}(u)}=( L^{[k]})^2\frac{\bar\phi_{(-1)}^{\,\prime}(u)^2}{2\bar\phi_{(-1)}(u)}
\end{align}
implies $\bar\phi_{(-1)}^{\,\prime}(u)=0$, namely $\bar\phi_{(-1)}(u)=\beta$, where $\beta$ is a non-zero constant. Along with \eqref{type0ti}, we have
\begin{align}
	\bar\phi_{(-1)}(u)=\beta,  \quad \bar\phi_{(1)}(u)=0, \quad  t^{[i]}(u)=\frac{u}{\beta}.
\end{align}
If $\beta<0$, an additional negative sign should be included in the above expression for $t^{[i]}(u)$.  We have already eliminated the translation degree of freedom of $t^{[i]}$ in the above expression. 
Turning our attention to the subleading order junction condition,
\begin{align}
	\sum_{i=0}^m\partial_\ell\phi^{[i]}_{(1)}(u)=-\sum_{i=0}^m L^{[i]}\frac{\bar\phi_{(-1)}^{\,\prime}(u)^2}{\bar\phi_{(-1)}(u)}=0=-\chi\bar\phi_{(1)}(u),
\end{align}
which is automatically satisfied. Furthermore, comparing with Eq.\eqref{phi(-1)u}, we find that $E=\alpha=0$. As a result, the continuity and  junction conditions for the extrinsic curvature and the dilaton are compatible.

In the scenario where all $m+1$ pages are of Type $0$, it is worth noting that the expression for $t^{[i]}(u)$ lacks translational symmetry with respect to $u$. This is because, the translational symmetry of $u$ is equivalent to that of $ t^{[i]}$ which have beed eliminated as a redundant degrees of freedom. Calculating the gravitational part of the action,
\begin{align}
	16\pi  G_{\mathrm N}\,I_{\mathscr V^{[i]}}=\frac{2\beta}{\epsilon^2L^{[i]}}\oint_\Sigma \mathrm du+\mathcal O(\epsilon^2).
\end{align}
We immediately observe that the Schwarzian derivative and the Schwarzian action are both equal to $0$, corresponding to an infinite inverse temperature.

Specifically, we examine the scenario in which $\mathscr P$ exhibits $\mathbb Z_{m+1}$ symmetry, with all $L^{[i]}=L$ and $\chi=\frac{m+1}{L}$. The continuity condition is automatically met in this context, and thus we directly compute the junction condition,
\begin{align}
	-(m+1)L\frac{\bar\phi_{(-1)}^{\,\prime}(u)^2}{\bar\phi_{(-1)}(u)}=(m+1)L\frac{\bar\phi_{(-1)}^{\,\prime}(u)^2}{2\bar\phi_{(-1)}(u)}\quad \Longrightarrow \quad \bar\phi_{(-1)}^{\,\prime}(u)=0.
\end{align}
As a result, we have $t(u)=L^2u/\beta$, where $\beta=\bar\phi_{(-1)}(u)$, and $E=\alpha=0$.

\subsubsection{All pages taking Type $-$}
We now consider all dilatons to be of Type $-$. It is worth emphasizing that we have selected a fixed coordinate system $t^{[i]}(u)$ for each page, ensuring that the continuity condition takes the form of Eq.\eqref{strongercontinu1}. Using Eq.\eqref{phi1(u)Type-}, the $p=1$ order continuity condition implies that
\begin{align}
	\notag
	  & \frac{( B^{[j]} L^{[j]})^2}{\bar\phi_{(-1)}(u)}\bigg\{\bigg[\tan\bigg( B^{[j]}\int \frac{\mathrm du}{\bar\phi_{(-1)}(u)}+ \theta^{[j]}\bigg)-\frac{\bar\phi_{(-1)}^{\,\prime}(u)}{ B^{[j]}}\bigg]^2-\frac{1}{2}\bigg(\frac{\bar\phi_{(-1)}^{\,\prime}(u)}{ B^{[j]}}\bigg)^2-1\bigg\} \\
	= & \frac{( B^{[k]} L^{[k]})^2}{\bar\phi_{(-1)}(u)}\bigg\{\bigg[\tan\bigg( B^{[k]}\int \frac{\mathrm du}{\bar\phi_{(-1)}(u)}+ \theta^{[k]}\bigg)-\frac{\bar\phi_{(-1)}^{\,\prime}(u)}{ B^{[k]}}\bigg]^2-\frac{1}{2}\bigg(\frac{\bar\phi_{(-1)}^{\,\prime}(u)}{ B^{[k]}}\bigg)^2-1\bigg\}
\end{align}
holds for any two indices $[j]$ and $[k]$. The manifestation of the continuity condition is such notably intricate that  a practical analytical expression for $\bar\phi_{(-1)}(u)$ cannot be derived. Hence, we must temporarily assume the validity of the continuity condition and then proceed from the junction condition. Using Eq.\eqref{phi1(u)Type-} again,
\begin{align}
	\notag
	 & \chi\bar\phi_{(1)}(u)=\sum_{i=0}^m\frac{\bar\phi_{(1)}(u)}{L^{[i]}}                                                                                                                                                                                                           \\
	 & =\sum_{i=0}^m\frac{ L^{[i]}( B^{[i]})^2}{\bar\phi_{(-1)}(u)}\bigg\{\bigg[\tan\bigg( B^{[i]}\int \frac{\mathrm du}{\bar\phi_{(-1)}(u)}\bigg)-\frac{\bar\phi_{(-1)}^{\,\prime}(u)}{ B^{[i]}}\bigg]^2-\frac{1}{2}\bigg(\frac{\bar\phi_{(-1)}^{\,\prime}(u)}{ B^{[i]}}\bigg)^2-1\bigg\},
\end{align}
in conjunction with Eq.\eqref{pdellphii(u)forType-}, the subleading-order  junction condition yields
\begin{align}
	0=\chi & \bar\phi_{(1)}(u)+\sum_{i=0}^m\partial_\ell\phi^{[i]}_{(1)}(u)=-\sum_{i=0}^m\frac{L^{[i]}}{2\bar\phi_{(-1)}(u)}\big(\bar\phi_{(-1)}^{\,\prime}(u)^2+4( B^{[i]})^2\big).
\end{align}
It can be observed that, even for Type $-$, the junction condition at $p=1$ order still depend solely on $\bar\phi_{(-1)}^{\,\prime}(u)$ rather than $\int \frac{\mathrm du}{\bar\phi_{(-1)}(u)}$, once again demonstrating its coordinate-independence (up to $\mathcal O(\epsilon^2)$). However, the above equation cannot hold, even when $\mathscr P$ possesses $\mathbb Z_{m+1}$ symmetry, due to the fact that $\bar\phi_{(-1)}^{\,\prime}(u)^2+4( B^{[i]})^2>0$.  Consequently, under any circumstances, it is impossible to glue together $m+1$ Type $-$ pages. We can see that $4( B^{[i]})^2$ is precisely the constant $ D^{[i]}$ that we previously defined, and $ D^{[i]}>0$ represents repulsion action. When all dilatons repel each other, the pages are inevitably unable to be joined together.

\subsubsection{General case: any combination of the three types}
Now, let us consider the most general scenario, which involves the possibility of combining pages of the three types in arbitrary proportions. We assume there are $m_+$ Type $+$ pages, $m_0$ Type $0$ pages, and $m_-$ Type $-$ pages, such that $m_++m_0+m_-=m+1$. The continuity condition at $p=1$ order is described by Eq.\eqref{phi1(u)allType} and \eqref{phi1(u)Type-}. We can see that the various combinations of types result in a complex form of the continuity condition. We are uncertain whether the continuity condition at $p=1$ order can hold in this situation, let alone provide analytical constraints on $\bar\phi_{(-1)}(u)$. Therefore, we directly examine the junction condition at $p=1$ order.

By eqs.\eqref{pdellphii(u)forallType}\eqref{pdellphii(u)forType-}, and eqs.\eqref{phi1(u)allType}\eqref{phi1(u)Type-}, we have
\begin{align}
	\label{alltypeO1junction}
	\notag
	&0=\chi\bar\phi_{(1)}(u)+\sum_{i=0}^m\partial_\ell\phi^{[i]}_{(1)}(u)=\sum_{i=0}^m \frac{1}{L^{[i]}}\bigg\{\bar\phi_{(1)}(u)+L^{[i]}\cdot\partial_\ell\phi^{[i]}_{(1)}(u)\bigg\}\\
	\notag
	&=\sum_j\frac{(A^{[j]})^2/L^{[j]}-L^{[j]}\,\bar\phi_{(-1)}^{\,\prime}(u)^2}{2\bar\phi_{(-1)}(u)}-\sum_k\frac{L^{[k]}\,\bar\phi_{(-1)}^{\,\prime}(u)^2}{2\bar\phi_{(-1)}(u)}-\sum_l\frac{L^{[l]}\bigl(\bar\phi_{(-1)}^{\,\prime}(u)^2+4(B^{[l]})^2\bigr)}{2\bar\phi_{(-1)}(u)}\\
	\notag
	&=-\frac{\bar\phi_{(-1)}^{\,\prime}(u)^2}{2\bar\phi_{(-1)}(u)}\bigg(\sum_j L^{[j]}+\sum_kL^{[k]}+\sum_lL^{[l]}\bigg)-\frac{1}{2\bar\phi_{(-1)}(u)}\bigg(\sum_j D^{[j]}L^{[j]}+\sum_lD^{[l]}L^{[l]}\bigg)\\
	&=-\frac{1}{2\bar\phi_{(-1)}(u)}\bigg(\sum_{i=0}^mL^{[i]}\bigg)\big(\bar\phi_{(-1)}^{\,\prime}(u)^2+E\big),
\end{align}
where, the summation $\sum_i$ is over all pages; $\sum_j$ is over all Type $+$ pages; $\sum_k$ over all Type $0$ pages; and $\sum_l$ over all Type $-$ pages. It is evident that the junction condition is equivalent to
\begin{align}
	\label{balancecondition}
	\bar\phi_{(-1)}^{\,\prime}(u)^2+E=0,
\end{align}
which proves that $\bar\phi_{(-1)}^{\,\prime}(u)$ can only be constant, denoted as $\alpha$, then
\begin{align}
	\bar\phi_{(-1)}(u)=\alpha u+\beta,\qquad E+\alpha^2=0.
\end{align}
The above equation is fully compatible with Eq.\eqref{phi(-1)u}, and demonstrates the self-consistency of the junction conditions. Now, we can provide a detailed description of the physical significance of the constants $ D^{[i]}$ and $E$ introduced in eqs.\eqref{Di} and \eqref{E}. The constant $ D^{[i]}$ represents the average contribution of the dilatons on each page $\mathscr V^{[i]}$, and $E$ is the average of $ D^{[i]}$. The role of $E$ for dilatons is analogous to that of tension $\chi$ for extrinsic curvature. Just as $\Sigma$ must have tension to maintain balance with the extrinsic curvature of the joined pages, the existence of $E$ is necessary to balance the dilatons, and the ``equilibrium condition'' is $E+\alpha^2=0$, with $\alpha$ exhibits the characteristic scale of the dilaton at the boundary. In particular, $E$ comprises contributions from all $ D^{[i]}$, and specially for Type $+$, $ D^{[i]}$ is less than $0$. The larger the proportion of Type $+$ pages, the larger the value of $\alpha$, reflecting an ``attractive'' action of Type $+$ dilatons which agglomerate all dilatons closely together. Hence, the greater the number of Type $+$ pages, the more robustly joined they become. For Type $0$, $ D^{[i]}$ is $0$, contributing nothing to $\alpha$, representing a neutral effect with neither attraction nor repulsion. For Type $-$, $ D^{[i]}$ is greater than $0$, and to some extent, the more Type $-$ pages there are, the smaller $\alpha$ becomes. If there are too many Type $-$ pages, it will prevent the attractive action of Type $+$ from maintaining the joining of the pages. Therefore, we can observe that it is the mutual exclusion action that prevents the joining of only Type $-$ pages.

When all the pages are of Type $+$ or all are of Type $0$, the continuity condition and junction condition independently yield the result $\bar\phi_{(-1)}^{\,\prime}(u)=\text{const}$. This strengthens our confidence in assuming the continuity condition Eq.\eqref{strongercontinu} which is stronger than Eq.\eqref{O(1)continu}. It can be noted that, at least for pages of Type $+$ and Type $0$, employing eqs.\eqref{classifydilaton+} and \eqref{classifydilaton0} to establish a specific coordinate system is reasonable. However, for pages of Type $-$, the approach of completely breaking the coordinate symmetry, eliminating the residual one-parameter degree of freedom in Eq.\eqref{classifydilaton-}, while using the strong assumption of continuity \eqref{strongercontinu}, appears less natural. Now, with the equilibrium condition $E+\alpha^2=0$ significantly simplifying the problem, we re-examine the continuity condition for Type $-$ at the $p=1$ order. Now, assuming that $\mathscr V^{[j]}$ is Type $-$, while $\mathscr V^{[k]}$ is either Type $0$ or Type $+$, then $\phi^{[j]}_{(1)}(u)=\phi^{[k]}_{(1)}(u)$ implies
\begin{align}
	\tan\bigg(\frac{ B^{[j]}}{\alpha}\ln \frac{u+\beta/\alpha}{L^{[j]}}+\theta^{[j]}\bigg)\quad \text{or}\quad \tan\bigg(\frac{B^{[j]}}{\beta}u+\theta^{[j]}\bigg)=f\big(\alpha,B^{[j]},A^{[k]},L^{[j]},L^{[k]}),
\end{align}
where the function $f\big(\alpha,  B^{[j]}, A^{[k]}, L^{[j]}, L^{[k]})$ is independent of $u$. However, the above equation cannot hold in a neighborhood of $u$ unless $ B^{[j]}=0$ (corresponding to a trivial dilaton).

This result may suggest that, for the pages $\mathscr V^{[i]}$ of Type $-$, there might not exist any coordinate system $t^{[i]}(u)$ such that $\phi^{[i]}_{(1)}(u)=\bar\phi_{(1)}(u)+\mathcal O(\epsilon^2)$ holds. In other words, for a finite cut-off, Type $-$ cannot be continuously joined with other types at the $p=1$ order. However, the leading-order continuity condition $\phi^{[i]}_{(-1)}(u)=\bar\phi_{(-1)}(u)+\mathcal O(\epsilon^2)$ still holds significance for Type $-$. If $\epsilon$ is sufficiently small, the leading-order continuity condition for Type $-$ is already enough. In this scenario, the junction condition at $p=1$ order does not prevent a small number of Type $-$ pages from gluing to pages of other types.

Finally, we will present a concrete example of joining pages with different types, where both the continuity and the junction conditions at subleading order are satisfied. Therefore, it is necessary that $m_- = 0$.

\subsubsection{Combination of Type $+$ and Type $0$}
With $m_- = 0$, we can express $m_++m_0=m+1$, where $m_+>0, m_0>0$. Let us arbitrarily designate $\mathscr V^{[j]}$ as Type $+$ and, at the same time, choose $\mathscr V^{[k]}$ as Type $0$. As a result, the continuity condition provides
\begin{align}
	\bar\phi_{(1)}(u)=\frac{\big( A^{[j]}- L^{[j]}\bar\phi_{(-1)}^{\,\prime}(u)\big)^2}{2\bar\phi_{(-1)}(u)}=\frac{\big( L^{[k]}\bar\phi_{(-1)}^{\,\prime}(u)\big)^2}{2\bar\phi_{(-1)}(u)}.
\end{align}
It can be solved immediately that $\bar\phi_{(-1)}^{\,\prime}(u) = \frac{ A^{[j]}}{ L^{[j]}+ L^{[k]}}$ or $\frac{ A^{[j]}}{ L^{[j]}- L^{[k]}}$, so $\bar\phi_{(-1)}^{\,\prime}(u)=\alpha$ is a constant. We assert that $\alpha\not=0$; otherwise, the above equation immediately leads to $ A^{[j]}=0$ for all Type $+$ pages, which would render the dilaton trivial, contradicting the premise. Then, let us choose both $\mathscr V^{[k]}$ and $\mathscr V^{[l]}$ to be Type $0$, the continuity condition gives
\begin{align}
	\bar\phi_{(1)}(u)=\frac{\big( L^{[k]}\bar\phi_{(-1)}^{\,\prime}(u)\big)^2}{2\bar\phi_{(-1)}(u)}=\frac{\big( L^{[l]}\bar\phi_{(-1)}^{\,\prime}(u)\big)^2}{2\bar\phi_{(-1)}(u)}.
\end{align}
This signifies that it is necessary for $ L^{[k]}= L^{[l]}$ to hold for any two Type $0$ pages, resulting in all Type $0$ pages sharing an equal AdS radius denoted as $L_0$. Now, it is obtained that $\big( A^{[j]}-\alpha L^{[j]}\big)^2=(\alpha L_0)^2$ for all Type $+$ pages $\mathscr V^{[j]}$. Hence, we further assume that there exist $n_1$ Type $+$ pages that satisfy $ A^{[j]}-\alpha  L^{[j]}=\alpha L_0$; consequently, there are $n_2 = m_+ - n_1$ pages that satisfy $ A^{[j]}-\alpha  L^{[j]}=-\alpha L_0$. We have obtained
\begin{align}
	\bar\phi_{(-1)}(u)=\alpha(u+\gamma), \qquad \bar\phi_{(1)}(u)=\frac{\alpha (L_0)^2}{2(u+\gamma)},
\end{align}
which can be used to solve the $t^{[i]}(u)$ of all pages. There are a total of three categories,
\begin{align}
	\notag
	 &  t^{[j]}(u)= L^{[j]}\bigg(\frac{u+\gamma}{ L^{[j]}}\bigg)^{1+ L_0/ L^{[j]}},\qquad   t^{[l]}(u)=\pm L^{[l]}\bigg(\frac{u+\gamma}{ L^{[l]}}\bigg)^{1- L_0/ L^{[l]}}; \\
	 &  t^{[k]}(u)=\pm\frac{1}{\alpha} \ln \frac{u+\gamma}{L_0},
\end{align}
where $t^{[j]}(u), t^{[l]}(u)$ represent the parameter equations of $\Sigma$ in the Type $+$ pages, satisfying the conditions $ A^{[j]}=\alpha( L^{[j]}+L_0)$ and $ A^{[l]}=\alpha( L^{[l]}-L_0)$ respectively. Additionally, $ t^{[l]}(u)$ takes the negative sign when $1-L_0/ L^{[l]}<0$, and the positive sign otherwise. As for $ t^{[k]}(u)$, it represents the equations of $\Sigma$ in Type $0$ pages, and it takes the negative sign when $\alpha<0$, and the positive sign otherwise.

Now, we calculate the subleading-order junction condition, which is equivalent to
\begin{align}
	\notag
	0 & =E+\alpha^2=\alpha^2-\biggl(m_0L_0+\sum_j  L^{[j]}\biggr)^{-1}\biggl(\sum_k\frac{\alpha^2}{ L^{[k]}}\bigl( L^{[k]}+L_0\bigr)^2+\sum_l\frac{\alpha^2}{ L^{[l]}}\bigl( L^{[l]}-L_0\bigr)^2\biggr) \\
	  & =-\alpha^2L_0\biggl(m_0L_0+\sum_j  L^{[j]}\biggr)^{-1}\bigg(\chi L_0+2\big(n_1-n_2-m_0\big)\bigg).
\end{align}
In which, $n_1+n_2=m_+$, $n_1+n_2+m_0=m+1$; the $\sum_j$ represents the sum over all Type $+$ pages; the $\sum_k$ sums over all Type $+$ pages satisfying the condition $ A^{[k]}=\alpha( L^{[k]}+L_0)$; and the $\sum_l$ sums over all Type $+$ pages satisfying $ A^{[l]}=\alpha( L^{[l]}-L_0)$. Consequently, the junction condition leads to
\begin{align}
	\chi L_0+2\big(n_1-n_2-m_0\big)=0.
\end{align}
If we set $\zeta \coloneqq L_0, m_0 = 0$, then the above equation simplifies to the junction condition \eqref{alltype+junction} where all the pages are of Type $+$. It is worth noting that the junction condition is entirely dependent on the number of each type and its AdS radius. Particularly, when $n_1\geqslant\frac{m+1}{2}$, the junction condition will inevitably not be satisfied, indicating that the number of pages satisfying $ A^{[k]}=\alpha( L^{[k]}+L_0)$ is less than half of the total. The ratio of the harmonic mean value of all Type $+$ AdS radii to $L_0$ must be a rational number.

At last, let us calculate the gravitational part of the action for each page,
\begin{subequations}
	\begin{align}
		 & 16\pi G_{\mathrm N}\,I_{\mathscr V^{[i]}}=\frac{2\alpha}{\epsilon^2L^{[i]}}\oint_\Sigma u\,\mathrm du+\frac{2\alpha L_0(L_0\pm L^{[i]})}{L^{[i]}}\oint_\Sigma\frac{1}{u}\,\mathrm du+\mathcal O(\epsilon^2), \\
		 & 16\pi G_{\mathrm N}\,I_{\mathscr V^{[j]}}=\frac{2\alpha}{\epsilon^2 L_0}\oint_\Sigma u\,\mathrm du+\mathcal O(\epsilon^2),
	\end{align}
\end{subequations}
where, page $\mathscr V^{[i]}$ is of Type $+$, and page $\mathscr V^{[j]}$ is of Type $0$. The $\mathcal{O}(1)$ term in the equation corresponds to the Schwarzian action, which is $0$ for each Type $0$ page. The value of the action is still dominated by $\alpha$.

\subsection{Brief summary of this section} 
We have now jointly solved all continuity and junction conditions for the dilaton, as well as the junction conditions for the extrinsic curvature, at the subleading order of accuracy. This has provided constraints on the basic parameters of the pages, which are very loose and easily satisfied. We consider all possible combinations of different types of pages, glued together in various proportions. The mutual gluing of Type $-$ pages is excluded, as their repulsive effects prevent such structures from satisfying the junction conditions at the subleading order. Similarly, the gluing of Type $-$ pages with the other two types is also excluded, as it fails to meet the continuity conditions at this order of accuracy. Of course, ignoring the continuity condition at this order would introduce a wider range of possibilities. What we can confirm is that the junction conditions at the subleading order are physical, whereas the continuity conditions at this order arise from our strong assumptions. However, through various actual calculations---such as all pages being Type $+$; all pages being Type $0$; or some pages being Type $+$ and others Type $0$---all of which were derived from the subleading order continuity conditions, we obtained results that are consistent with the independent application of the junction conditions. This strongly suggests that the subleading order continuity assumption is reasonable.

However, it is neither necessary nor correct to extend this assumption and apply it to the continuity or even the junction conditions at higher orders. As we have seen, the physical constraints at the current subleading order determine the dilaton to have a single parameter, $\alpha$, which is physical and directly related to the dilaton's invariant, $A^2-4BC$, thus reflecting the characteristic scale of the dilaton. Higher-order junction conditions cannot, and should not, determine the value of $\alpha$, as this would be equivalent to fully determining the dilaton. Physically, the dilaton should be determined by boundary conditions, not junction conditions. If we force the assumption that higher-order conditions are physical in some coordinate $t^{[i]}$, solving them is likely to yield incompatible or inconsistent results.  

\section{Conclusions and outlook\label{sec6}}
By gluing together multiple bulk spacetimes along a common interface and imposing gravitational consistency at the junction, one can obtain the booklet structure and derive the multiway junction condition for it. To examine the physical constraints enforced by this condition in explicit setups, we construct and solve the multiway junction condition within the framework of JT gravity. Specifically, we consider a configuration consisting of $m+1$ AdS$_2$ spacetimes, each of which is endowed with JT gravity and is truncated along a closed curve near the asymptotic boundary. By identifying all these truncation curves, we define a common interface, referred to as the defect, along which the geometries are glued together. This construction gives rise to a multi-boundary geometry that is topologically distinct from that of the replica wormhole configuration. We assume that the defect possesses tension which is coupled to the dilaton. Since JT gravity involves both the dilaton field and the metric as dynamical variables, the corresponding junction conditions involve constraints on both the extrinsic curvature and the dilaton at the interface.

Since the junction condition is local and coordinate-independent, for convenience, we choose to solve it in Poincar\'e coordinates. Following the standard approach to JT gravity in the literature, we normalize the induced metric $h$ on the defect and introduce a parameter $\epsilon$ to characterize how close the truncation is to the asymptotic boundary. At this stage, $h\propto 1/\epsilon^2$. Expanding the constraint equations---namely, the junction conditions for both the extrinsic curvature and the dilaton, as well as the continuity condition for the dilaton across the defect---in powers of $\epsilon$, we solve them order by order. However, a significant difficulty arises: the series expansion procedure breaks the coordinate invariance of the constraint equations. In other words, each order of the expansion may depend on the specific choice of Poincar\'e coordinates. Through calculation, we find that the leading and subleading orders of the junction condition, as well as the leading order of the continuity condition, are coordinate-independent and thus physically meaningful. In contrast, higher orders no longer possess coordinate invariance.  If some higher-order terms yield conclusions that conflict with the aforementioned lower-order constraints, we can assert that those higher orders lack physical significance. Nevertheless, exceptions exist. For instance, the subleading order of the continuity condition is not coordinate-independent; yet explicit calculations demonstrate that this constraint is not only compatible with the lower-order results but also exhibits considerable naturalness and rationality. This constraint can fully determine the form of the dilaton and the defect, leaving no room for further freedom at higher orders.  In particular, the leading order results of the junction conditions for both the extrinsic curvature and the dilaton are trivial, consistent with conclusions in other literature. However, solving the constraints at the subleading order yields many nontrivial corrections, which only arise when gluing more than two bulks together, underscoring the significance of the booklet structure to some extent.

The general solution of the dilaton contains three parameters; however, these parameters actually involve redundancies, and the true physical degrees of freedom amount to only one. When solving the multiway junction conditions, working with the original form of the dilaton containing three parameters proves to be highly problematic. For this reason, we first attempt to eliminate the redundant degrees of freedom. The isometry group of the metric is generated by the Lie algebra $\mathfrak{sl}(2,\mathbb R)$, while the solution space of the dilaton forms a representation of the $\mathfrak{sl}(2,\mathbb R)$ algebra. By employing the Killing form of the $\mathfrak{sl}(2,\mathbb R)$ algebra, we construct an invariant on the dilaton solution space. This invariant naturally classifies all dilaton solutions into three inequivalent types. Furthermore, for each distinct type of dilaton, we fix it into a standard form via an isometric isomorphism transformation, where this standard form depends on only a single parameter representing the invariant we constructed. This procedure is equivalent to identifying a special class of reference frames for each type of dilaton. The different types of dilaton correspond to different signs of the potential, and thus, as a vivid qualitative characterization, we refer to the dilatons with negative, positive, and zero potentials as having attractive, repulsive, or neutral effects, respectively.

As previously mentioned, only the leading and subleading orders of the junction conditions for the dilaton and the extrinsic curvature carry physical significance. In contrast, for the continuity condition of the dilaton, only the leading order is necessarily physical. However, in concrete calculations, the junction and continuity conditions are closely intertwined. To ensure that the solutions to both sets of conditions possess the same level of precision, we make a physically well-motivated assumption: we impose the second-order constraints from the continuity condition. Since the coordinate invariance of the continuity condition is broken by the series expansion, it becomes important to determine in which Poincar\'e coordinate the second-order continuity condition should be considered. Fortunately, the classification of dilaton solutions and the subsequent fixation of each type into a standard form provide a natural resolution to this issue. As shown in the calculations and discussion in section \ref{subsec4E}, this assumption proves to be highly reasonable. We must consider all possible configurations involving $m+1$ bulks, where the dilaton in each bulk can be independently selected from among the three inequivalent types. Our analysis reveals that only when a sufficient number of these bulks carry a dilaton with attractive effect can the entire junction be maintained. The second-order continuity condition reflects the strength and stability of the junction. Moreover, when second-order continuity is taken into account, bulks with repulsive-type dilatons are strictly excluded --- a conclusion that holds independently of the choice of coordinates. As a result, the only allowed configurations are the following three cases: all bulks carry an attractive-type dilaton; all bulk regions carry a neutral-type dilaton; a specific combination of bulks carrying attractive and neutral-type dilatons in a precisely determined ratio. We have carried out a complete analysis for each of these cases and have determined the explicit form of the corresponding dilaton solutions.

Although JT gravity is a toy model, its excellent solvability and the low dimensionality of spacetime allow us to freely specify the shape of the cutoff curves to a large extent. This, in turn, imposes fewer constraints on solving the multiway junction conditions and grants greater generality to the analysis. Our next goal is to extend the study to higher-dimensional and physically more realistic models, and to solve their corresponding multiway junction conditions. Ultimately, we aim to demonstrate that the booklet structures can arise as saddle points in the gravitational path integral, analogous to replica wormholes, and contribute to the computation of entanglement entropy or other physical quantities via path integrals. Such a result would provide new perspectives and methods for addressing the black hole information paradox.
\appendix

\section{Change of coordinates in the series expansions}\label{changecoord}
Let us provide a detailed explanation of the $\epsilon$ expansion of $\sum_{i=0}^m K^{[i]}$. The extrinsic curvature $K^{[i]}$ of the truncated curve $\Sigma$ can be computed using the parametric equations $(t^{[i]},z^{[i]})$ in Poincar\'e coordinates, where $z^{[i]}$ is determined by the differential equation \eqref{SigmaParaEq} and is thus a function of $t^{[i]}$ and $\epsilon$. Consequently, the dependence of $K^{[i]}$ on $t^{[i]}$ and $\epsilon$ can be denoted by $\mathscr F[t^{[i]},\epsilon]$. Formally,
\begin{align}
	\label{formalexpand}
	-\chi=\sum_{i=0}^m K^{[i]}=\sum_{i=0}^m \mathscr F[t^{[i]},\epsilon]=\sum_{p=0}^\infty\sum_{i=0}^m  \frac{1}{p!}\left(\frac{\partial}{\partial\epsilon}\right)^p\mathscr F[t^{[i]},0]\epsilon^p.
\end{align}
Let us clarify the notation. The expression $\mathscr F[t^{[i]},\epsilon]$ serves as a shorthand notation depending on not only $t^{[i]}$ but also its derivatives, such as $t^{[i]\prime},t^{[i]\prime\prime},\cdots$, and potentially other independent variables
\footnote{
	For example, the parameters $A^{[i]}, B^{[i]}, C^{[i]}$ required to determine the dilaton in the following sections can be considered as such independent variables.
	} 
that depend on $t^{[i]}$. Thus, when we write $\frac{\partial\mathscr F}{\partial t^{[i]}}$, this denotes an ordered set of functions, each representing the partial derivative with respect to one of the independent variables associated with $t^{[i]}$.

Each order of this formal series \eqref{formalexpand} corresponds to the same order in \eqref{seriesjunctionK}, and this approach to expansion also applies to the dilaton discussed later. Notably, our initial choice of $t^{[i]}$ is arbitrary, meaning that the form of $\mathscr F$ is independent of the choice of Poincar\'e coordinates, i.e., $\mathscr F[t^{[i]},\epsilon]=\mathscr F[\tilde t^{[i]},\epsilon]$, where $t^{[i]}\to\tilde t^{[i]}$ is an isometric coordinate transformation. Furthermore, the form of each order $\partial_\epsilon^p\mathscr F$ in \eqref{seriesjunctionK} remains invariant under coordinate transformations. However, the value of each order in the expansion, $\partial_\epsilon^p\mathscr F[t^{[i]},0]$, may depend on the specific choice of $t^{[i]}$. In other words,
\begin{align}
	\label{distinctorder}
	\left(\frac{\partial}{\partial\epsilon}\right)^p\mathscr F[t^{[i]},0]\not=\left(\frac{\partial}{\partial\epsilon}\right)^p\mathscr F[\tilde t^{[i]},0].
\end{align}
The reasoning is as follows: as discussed in Appendix \ref{appx2}, the new coordinates $\tilde t^{[i]}$ may depend on both $t^{[i]}$ and $z^{[i]}$, which we denote by $\tilde t^{[i]}=\mathscr H[t^{[i]},\epsilon]$, where the form of $\mathscr{H}$ depends on certain parameters involved in the specific metric-preserving transformation. Thus,
\begin{align}
	\label{FoH}
	\mathscr F[t^{[i]},\epsilon]=\mathscr F[\mathscr H[t^{[i]},\epsilon],\epsilon],
\end{align}
Physically, if $\epsilon$ is perturbed by $\epsilon\to \epsilon+\delta\epsilon$, then the parameterization $t^{[i]}(u)$ will correspond to a new curve $\Sigma^\prime$, and the coordinate transformation for $t^{[i]}(u)$ should be $\mathscr H[t^{[i]},\epsilon+\delta\epsilon]$ rather than $\tilde t^{[i]}=\mathscr H[t^{[i]},\epsilon]$, meaning that $\tilde t^{[i]}$ will correspond to yet another new curve $\Sigma^{\prime\prime}$. Therefore,
\begin{align}
	\mathscr F[t^{[i]},\epsilon+\delta\epsilon]=\mathscr F[\mathscr H[t^{[i]},\epsilon+\delta\epsilon],\epsilon+\delta\epsilon]\not=\mathscr F[\mathscr H[t^{[i]},\epsilon],\epsilon+\delta\epsilon]=\mathscr F[\tilde t^{[i]},\epsilon+\delta\epsilon].
\end{align}
The above expression actually determines the derivatives of various orders of $\mathscr F$, which can directly prove that, although \eqref{FoH} holds for any $\epsilon$, it is still possible to derive \eqref{distinctorder}. Furthermore, we can examine in detail the quantitative relationships between expansion terms in different coordinates. First, we differentiate both sides of eq.\eqref{FoH} with respect to $\epsilon$,  
\begin{align}
	\label{DF=DFoH}
	\notag
	\left(\frac{\partial}{\partial\epsilon}\right)^p\mathscr F[t^{[i]},0]&=\left(\frac{\partial\mathscr H}{\partial\epsilon}[t^{[i]},0]\cdot\frac{\partial}{\partial\tilde t^{[i]}}+\frac{\partial}{\partial\epsilon}\right)^p\mathscr F[\mathscr H[t^{[i]},0],0]\\
	&=\sum_{\Bbbk=0}^p\frac{p!}{\Bbbk!\,(p-\Bbbk)!}\left(\frac{\partial\mathscr H}{\partial\epsilon}[t^{[i]},0]\cdot\frac{\partial}{\partial\tilde t^{[i]}}\right)^{p-\Bbbk}\left(\frac{\partial}{\partial\epsilon}\right)^\Bbbk\mathscr F[\mathscr H[t^{[i]},0],0].
\end{align}
We will further clarify the notation. As per previous conventions, $\frac{\partial}{\partial\tilde t^{[i]}}$ denotes the partial derivative over all expressions involving $\tilde t^{[i]}$ (or equivalently, $\mathscr H$), resulting in an ordered set of partial derivative functions. Similarly, $\frac{\partial\mathscr H}{\partial\epsilon}$ refers to the derivative with respect to $\epsilon$ for all quantities that depend on $\tilde t^{[i]}$ (or $\mathscr H$). Therefore, $\frac{\partial\mathscr H}{\partial\epsilon}[t^{[i]},0]\cdot\frac{\partial}{\partial\tilde t^{[i]}}$ represents the dot product of two ordered sets of functions. On the other hand, the isolated operator $\frac{\partial}{\partial\epsilon}$ takes the partial derivative with respect to $\epsilon$, while keeping all quantities involving $\tilde t^{[i]}$ (or $\mathscr H$) unchanged, meaning this operator does not affect the function $\frac{\partial\mathscr H}{\partial\epsilon}$. Consequently, the two operators, $\frac{\partial}{\partial\epsilon}$ and $\frac{\partial\mathscr H}{\partial\epsilon}[t^{[i]},0]\cdot\frac{\partial}{\partial\tilde t^{[i]}}$, commute, which allows us to derive the second equality of \eqref{DF=DFoH}. Now, starting from the formal expansion at $\tilde t^{[i]}$, 
\begin{align}
	\label{symmbreak}
	\notag
	&\sum_{p=0}^\infty\frac{1}{p!}\left(\frac{\partial}{\partial\epsilon}\right)^p\mathscr F[\mathscr H[t^{[i]},\epsilon],0]\epsilon^p
	=\sum_{p=0}^\infty\frac{1}{p!}\epsilon^p\sum_{\Bbbk=0}^\infty\frac{1}{\Bbbk!}\epsilon^\Bbbk\left(\frac{\partial\mathscr H}{\partial\epsilon}[t^{[i]},0]\cdot\frac{\partial}{\partial\tilde t^{[i]}}\right)^\Bbbk\left(\frac{\partial}{\partial\epsilon}\right)^p\mathscr F[\mathscr H[t^{[i]},0],0]\\
	&=\sum_{p=0}^\infty\frac{1}{p!}\epsilon^{p}\sum_{\Bbbk=0}^{p}\frac{p!}{\Bbbk!(p-\Bbbk)!}\left(\frac{\partial\mathscr H}{\partial\epsilon}[t^{[i]},0]\cdot\frac{\partial}{\partial\tilde t^{[i]}}\right)^{p-\Bbbk}\left(\frac{\partial}{\partial\epsilon}\right)^\Bbbk\mathscr F[\mathscr H[t^{[i]},0],0]
\end{align}
Using Eq.\eqref{DF=DFoH}, we can immediately transform the expression \eqref{symmbreak} into Eq.\eqref{formalexpand}, where in the second equality of \eqref{symmbreak}, we have performed a change of summation variables, $(p+\Bbbk,p)\to (p,\Bbbk)$. Therefore, Eq.\eqref{DF=DFoH} quantitatively demonstrates the influence of lower-order terms on higher-order terms under the coordinate transformation $t^{[i]}\to\tilde t^{[i]}$.

In eq.\eqref{phiuexpand}, our choice of coordinates $(t^{[i]}, z^{[i]})$ is arbitrary, so the dependence of $\phi^{[i]}_{(p)}$ on $t^{[i]}$ and $A^{[i]},B^{[i]},C^{[i]}$, denoted by $\mathscr F_{(p)}$, i.e., $\phi^{[i]}_{(p)}=\mathscr F_{(p)}\big[t^{[i]}, A^{[i]},B^{[i]},C^{[i]}\big]$\footnote{Similar to \eqref{formalexpand}, by taking 
$\mathscr F[t^{[i]},A^{[i]},B^{[i]},C^{[i]},\epsilon]\coloneqq\epsilon\phi^{[i]}\big|_\Sigma$, we have $\mathscr F_{(p)}=\frac{1}{(p+1)!}\left(\frac{\partial}{\partial\epsilon}\right)^{(p+1)} \mathscr F\big|_{\epsilon=0}$.}, remains unchanged. However, for different coordinates $t^{[i]}$ and $\tilde t^{[i]}$, the specific value of $\phi^{[i]}_{(p)}$ may vary, meaning that it is possible to have
\begin{align}
	\mathscr F_{(p)}\left[t^{[i]}, A^{[i]},B^{[i]},C^{[i]}\right] \not= \mathscr F_{(p)}\left[\tilde t^{[i]},\tilde A^{[i]},\tilde B^{[i]},\tilde C^{[i]}\right].
\end{align}
The two sets of coordinates $t^{[i]}$ and $\tilde t^{[i]}$ differ by a metric-preserving  diffeomorphism $\exp(s\xi)$. 
Using Iwasawa decomposition~\cite{Deitmar2009}, each metric-preserving transformation can be expressed as a product in the form of $\exp\big(\theta(\xi_b/L^{[i]}+L^{[i]}\xi_c)\big)^*\exp(a\xi_a)^*\exp(c\xi_c)^*$. Where, the actions of $\exp(a\xi_a)^*$ and $\exp(c\xi_c)^*$ on $ t^{[i]}(u)$ result in only an $\mathcal O(1)$ correction without $\epsilon$. Therefore, when $\exp(a\xi_a)^*$ and $\exp(c\xi_c)^*$ act on $\phi^{[i]}\big|_\Sigma$, the expansion terms at each order transform independently, without mutual influence. Consequently, for any $p$, it is certain that the value of $\phi^{[i]}_{(p)}(u)$ remains invariant under the action of $t^{[i]}\to \exp(a\xi_a)^*\exp(c\xi_c)^*t^{[i]}$. 

However, for the remaining $\exp\big(\theta(\xi_b/L^{[i]}+L^{[i]}\xi_c)\big)^*$, the setuation is different. The value of $\exp\big(\theta(\xi_b/L^{[i]}+L^{[i]}\xi_c)\big)^*t^{[i]}$ depends on both $ t^{[i]}(u)$ and $ z^{[i]}(u)$, leading to $\mathcal O(\epsilon^2)$ corrections in $ t^{[i]}(u)$.  Consequently, $\phi^{[i]}_{(p)}(u)$ also experience a $\mathcal O(\epsilon^2)$ correction: $\phi^{[i]}_{(p)}(u)\to \phi^{[i]}_{(p)}(u)+\psi(u)\epsilon^2$, leading to a corresponding $\mathcal O(1)$ correction in $\phi^{[i]}_{(p+2)}(u)$ term: $\phi^{[i]}_{(p+2)}(u)\to \phi^{[i]}_{(p+2)}(u)-\psi(u)$. Clearly, the behavior of lower-order terms has an impact on higher-order terms, such that the series expansion \eqref{phiuexpand} is not unique. This is reasonable: in expanding $\phi^{[i]}$, we used Eq.\eqref{epsilonO3}, treating $t^{[i]}(u)$ and $\epsilon$ as independent, essentially choosing a specific coordinate system.

Starting from any coordinate system $(t,z)$, by adjusting the parameters $\theta,a,c$, one can obtain all Poincar\'e coordinates through the action $\exp\big(\theta(\xi_b/L^{[i]}+L^{[i]}\xi_c)\big)^*\exp(a\xi_a)^*\\ \cdot\exp(c\xi_c)^*(t,z)$. Furthermore, $\exp(a\xi_a)^*\exp(c\xi_c)^*t^{[i]}$ and $t^{[i]}$ yield identical series expansions, so all elements of the form $\exp(a\xi_a)^*\exp(c\xi_c)^*$ form a (non-normal) subgroup, denoted $\mathrm{AN}\subset \mathrm{SL}(2,\mathbb R)$. This subgroup defines an equivalence class of coordinates, called the $\mathrm{AN}$-class. Meanwhile, $\exp\big(\theta(\xi_b/L^{[i]}+L^{[i]}\xi_c)\big)^*t^{[i]}$ produces a series expansion distinct from $t^{[i]}$, meaning each $\mathrm{AN}$-class can be labeled by a parameter $\theta$ (up to a translation in $\theta$)\footnote{This labeling is not unique due to the non-normal nature of the $\mathrm{AN}$ group, even though we treat labels differing by a translation in $\theta$ as equivalent.}. Next, when considering the gluing of $m+1$ pages along $\Sigma$, different pages may carry distinct types of dilatons. It seems necessary to select an $\mathrm{AN}$-equivalence class of Poincar\'e coordinates for each page. Although the junction condition itself is coordinate-independent, this choice becomes somewhat arbitrary. However, since practical calculations involve series expansions, symmetry breaking is induced in the choice of coordinates. Nevertheless, the constraints given by the leading-order terms are physically meaningful. For the transformation $t^{[i]}\to \tilde t^{[i]}=\mathscr H[t^{[i]},\epsilon]$, the new parameters $\tilde A^{[i]},\tilde B^{[i]},\tilde C^{[i]}$ must depend solely on $A^{[i]},B^{[i]},C^{[i]}$ and be independent of $\epsilon$ and $t^{[i]}$ (the derivation in section \ref{sec3} confirms this). Eq.\eqref{DF=DFoH} gives us, 
\begin{align}
	\label{partialF}
	\notag
	&\mathscr F_{(-1)}\big[t^{[i]},A^{[i]},B^{[i]},C^{[i]}\big]=\mathscr F_{(-1)}\big[\mathscr H[t^{[i]},0],\tilde A^{[i]},\tilde B^{[i]},\tilde C^{[i]}\big]\\
	&=\mathscr F_{(-1)}\big[\tilde t^{[i]},\tilde A^{[i]},\cdots\big]-\frac{\partial\mathscr H}{\partial\epsilon}[t^{[i]},0]\cdot\frac{\partial}{\partial \tilde t^{[i]}}\mathscr F_{(-1)}\big[\mathscr H[t^{[i]},0],\tilde A^{[i]},\cdots\big]\epsilon+\mathcal O(\epsilon^2).
\end{align}
According to \eqref{SL2Rgroup} in appendix \ref{appx2}, we have $\frac{\partial\mathscr H}{\partial \epsilon}[t^{[i]},0]\equiv 0$. Thus, if we replace $\mathscr F_{(-1)}\big[t^{[i]},\cdots\big]$ in a given equation with $\mathscr F_{(-1)}\big[\tilde t^{[i]},\cdots\big]$, this substitution will at most introduce corrections of $\mathcal O(\epsilon^2)$, which will not fundamentally alter the original equation, or alternatively, we can simply take $\epsilon\to 0$, so the original equation remains valid. Continuing with higher-order terms, according to \eqref{DF=DFoH}, under coordinate transformations, each order of expansion is influenced solely by lower-order terms and, in turn, only affects higher-order terms. Fortunately, for the junction conditions of the external curvature \eqref{seriesjunctionK}, we have already observed that the leading order does not actually depend on $t^{[i]}$, so the subleading order term (excluding vanishing orders) of \eqref{seriesjunctionK} is also physical. Through subsequent calculations, we will see that this property also applies to the junction conditions for the dilaton. Therefore, in this paper, we will solve the two junction conditions up to the subleading order. 

\section{General solution of the dilaton\label{appx1}}
In this appendix, we provide a detailed process for solving the equations of motion for the dilaton given in \eqref{motiondilaton}, which is equivalent to proving the expression in \eqref{solutionofdilaton}. First, we expand \eqref{motiondilaton} into its component form in the Poincar\'e coordinate basis,
\begin{subequations}
	\begin{align}
		\label{1}
		&\phi^{[i]}-z\partial_z\phi^{[i]}-z^2\partial^2_z\phi^{[i]}=0,\\
		\label{2}
		&\partial_t\big(\phi^{[i]}+z\partial_z\phi^{[i]}\big)=0,\\
		\label{3}
		&\phi^{[i]}+z\partial_z\phi^{[i]}+z^2\partial^2_t\phi^{[i]}=0.
	\end{align}
\end{subequations}
Let $\partial_t$ act on eq.\eqref{3} and subtract eq.\eqref{2} from it, 
\begin{align}
	\partial_t^3\phi^{[i]}=0 \quad \Longrightarrow \quad \phi^{[i]}(t,z)=f(z)t^2+g(z)t+h(z),
\end{align}
where the undetermined functions $f(z),g(z),h(z)$ arise as ``constants of integration'' (independent of the variable $t$). Note that eq.\eqref{2} implies that $\phi^{[i]}+z\partial_z\phi^{[i]}$ is independent of the coordinate $t$, which means that the following expression 
\begin{align}
	\phi^{[i]}+z\partial_z\phi^{[i]}=\partial_z(z\phi^{[i]})=\big(zf(z)\big)'t^2+\big(zg(z)\big)'t+\big(zh(z)\big)'
\end{align}
is actually independent of $t$, and thus,
\begin{align}
	(zf(z))'=0, \qquad (zg(z))'=0.
\end{align}
Introduce two constants of integration, 
\begin{align}
	g(z)=\frac{A}{z},\quad f(z)=\frac{B}{z},
\end{align}
Using eq.\eqref{3} again, we have
\begin{align}
    \label{zhprime}
	(zh(z))'=\phi^{[i]}+z\partial_z\phi^{[i]}=-z^2\partial_t^2\phi^{[i]}=-2f(z)z^2=-2Bz.
\end{align}
Solving the above equation gives
\begin{align}
    \label{h(z)}
	h(z)=\frac{-Bz^2+C}{z},
\end{align}
where $C$ is a constant of integration. We have now effectively determined the form of the dilaton solution. However, note that we have not yet used eq.\eqref{1}. To check the compatibility of the above solution with \eqref{1}, we will rewrite \eqref{1} in the following form:
\begin{align}
	\phi^{[i]}+z\partial_z\phi^{[i]}=z\partial_z\big(\phi^{[i]}+z\partial_z\phi^{[i]}\big),\quad \Longrightarrow \quad \bigg(\frac{\phi^{[i]}+z\partial_z\phi^{[i]}}{z}\bigg)'=0.
\end{align}
Substituting \eqref{zhprime}, it is easy to verify that the above expression holds, 
\begin{align}
	\bigg(\frac{\phi^{[i]}+z\partial_z\phi^{[i]}}{z}\bigg)'=\bigg(\frac{(zh(z))'}{z}\bigg)'=\frac{\mathrm d}{\mathrm dz}(-2B)=0, 
\end{align}
which demonstrates the compatibility of \eqref{h(z)} with \eqref{1}. Now, the form of the solution to the dilaton's equation of motion can be written as \eqref{solutionofdilaton}. 

Eq.\eqref{solutionofdilaton} is derived from the equation of motion \eqref{motiondilaton}, making it a necessary condition for \eqref{motiondilaton}. Conversely, substituting \eqref{solutionofdilaton} into the equation of motion \eqref{motiondilaton} demonstrates its sufficiency. Thus, $\phi^{[i]}$ is a solution to the equation of motion if and only if it takes the form given in \eqref{solutionofdilaton}.

\section{Some results for the Lie algebra $\mathfrak{sl}(2,\mathbb R)$\label{appx2}}
Given any generator $\xi$ in the Killing vector space,
\begin{align}
    \xi=a\xi_a+b\xi_b+c\xi_c,
\end{align}
where $\xi_a,\xi_b,\xi_c$ are defined by eq.\eqref{generatorofsl2R}. The coefficient $a$ is dimensionless; $b$ share the same dimensionality with $1/ L^{[i]}$; and $c$ have identical dimensionality with $ L^{[i]}$. Our goal is to solve the specific form of the one-parameter metric-preserving  diffeomorphism $\exp(s\xi)$, with $s$ representing a dimensionless parameter. Let $\exp(s\xi)^*(t,z)=\big(t(s),z(s)\big)$, where $t(0)=t, z(0)=z$. At any $s$, the tangent vector of the integral curve corresponds to the local value of $\xi$, thus
\begin{align}
    \label{tz}
    t^\prime(s)=at(s)+b\big(t(s)^2+z(s)^2\big)+c,\quad z^\prime(s)=az(s)+2bt(s)z(s).
\end{align}
If $b=0$, by employing perturbation methods, we can treat $b$ as a small quantity, and taking the limit $b\to 0$ at last. In fact, if we solve directly without employing perturbation methods, the cases where $b$ is zero and non-zero exhibit formal consistency, which naturally results from continuity. 
 Therefore, we can assume $b\not=0$, then
\begin{align}
    \label{tspmzs}
    \notag
	t^\prime(s)\pm z^\prime(s)=&b\big(t(s)\pm z(s)\big)^2+a\big(t(s)\pm z(s)\big)+c\\
    =&b\bigl(t(s)\pm z(s)-\varpi_0\bigr)\bigl(t(s)\pm z(s)-\varpi_1\bigr),
\end{align}
where $\varpi_0,\varpi_1$ are defined as
\begin{align}
    \varpi_0=\frac{-a+\sqrt{a^2-4bc}}{2b}, \quad \varpi_1=\frac{-a-\sqrt{a^2-4bc}}{2b}.
\end{align}
Certainly, these two roots may assume complex values. Therefore, the most suitable approach is to introduce a complex structure on the Poincar\'e patch of AdS$_2$, resulting in a Riemann surface, then the metric-perserving transformations will be the M\"obius transformations \cite{Jost2002}, resembling the form provided in eq.\eqref{tskappat}. However, for establishing the specific link between the transformations and the coefficients $a, b, c$, we undertake a detailed calculation. Assuming $\varpi_0\not=\varpi_1$, eq.\eqref{tspmzs} can be reformulated as
\begin{align}
    \frac{t^\prime(s)\pm z^\prime(s)}{t(s)\pm z(s)-\varpi_0}-\frac{t^\prime(s)\pm z^\prime(s)}{t(s)\pm z(s)-\varpi_1}=b(\varpi_0-\varpi_1),
\end{align}
the solution is
\begin{align}
    \frac{t(s)\pm z(s)-\varpi_0}{t(s)\pm z(s)-\varpi_1}=e^{b(\varpi_0-\varpi_1)\,s}\cdot \frac{t\pm z-\varpi_0}{t\pm z-\varpi_1},
\end{align}
hence we obtain the form of $\exp(s\xi)^*$,
\begin{align}
    \label{tskappat}
   \exp(s\xi)^*(t\pm z)= t(s)\pm z(s)=\frac{\kappa_{00}(s)\cdot\bigl(t\pm z\bigr)+\kappa_{01}(s)}{\kappa_{10}(s)\cdot\bigl(t\pm z\bigr)+\kappa_{11}(s)},
\end{align}
where the matrix elements of $\kappa(s)$ are defined as
\begin{align}
    \label{SL2Rmatrix}
    \kappa(s)=
    \left(\begin{matrix}
        \vspace{12pt}
        \dfrac{e^{-b(\varpi_0-\varpi_1)s/2}\varpi_0-e^{b(\varpi_0-\varpi_1)s/2}\varpi_1}{\varpi_0-\varpi_1}   &\dfrac{\varpi_0\varpi_1\big(e^{b(\varpi_0-\varpi_1)s/2}-e^{-b(\varpi_0-\varpi_1)s/2}\big)}{\varpi_0-\varpi_1}\\
    \dfrac{e^{-b(\varpi_0-\varpi_1)s/2}-e^{b(\varpi_0-\varpi_1)s/2}}{\varpi_0-\varpi_1}&\dfrac{e^{b(\varpi_0-\varpi_1)s/2}\varpi_0-e^{-b(\varpi_0-\varpi_1)s/2}\varpi_1}{\varpi_0-\varpi_1}
    \end{matrix}\right).
\end{align}
It is easy to verify that
\begin{align}
    \det \kappa(s)=\kappa_{00}(s)\kappa_{11}(s)-\kappa_{01}(s)\kappa_{10}(s)=1.
\end{align}
The group $\mathrm{SL}(2,\mathbb R)$ is originally defined as the set of all $2 \times 2$ real matrices with determinant equal to $1$. Consequently, if the matrix $\kappa(s)$ consists of real numbers, it is an element of the group $\mathrm{SL}(2,\mathbb R)$. In this case, eq.\eqref{tskappat} defines a representation/group homomorphism $\mathscr{R}$ of $\kappa(s)$, denoted as $\exp(s\xi)=\mathscr R\kappa(s)$, which is expressed as follows: for any point $q$ with coordinates $(t,z)$, the coordinates of the point $\mathscr R\kappa(s)(q)=\exp(s\xi)(q)$ are given by $\big(t(s),z(s)\big)$, as determined by eq.\eqref{tskappat}. It is straightforward to prove that $\mathscr R$ is well-defined, 
\begin{align}
    \label{grouphomo}
    \mathscr R\tilde\kappa(\tilde s)\circ \mathscr R\kappa(s)=\mathscr R\big[\tilde\kappa(\tilde s)\kappa(s)\big].
\end{align}
Specifically, we utilize the coordinate functions $x^\mu(q) = (t,z)$. Then \eqref{tskappat} can be expressed simply as $\exp(s\xi)^*x^\mu(q)=\mathscr F\left[\kappa(s),x^\mu(q)\right]$, where $\mathscr F$ is introduced to abbreviate the complex expression on the right-hand side of the equation. A direct calculation shows that $\mathscr F\left[\tilde\kappa,\mathscr F\left[\kappa,x^\mu\right]\right]=\mathscr F\left[\tilde\kappa\kappa,x^\mu\right]$, and noting that $\exp(s\xi)^*x^\mu(q)=x^\mu(\mathscr R\kappa(q))$, then
\begin{align}
    \notag
    x^\mu\big(\mathscr R\tilde\kappa\circ\mathscr R\kappa(q)\big)&=\mathscr F\left[\tilde\kappa,x^\mu\big(\mathscr R\kappa(q)\big)\right]=\mathscr F\left[\tilde\kappa,\mathscr F[\kappa,x^\mu]\right]=\mathscr F\left[\tilde\kappa\kappa,x^\mu\right]\\
    &=x^\mu\big(\mathscr R\left[\tilde\kappa\kappa\right](q)\big).
\end{align}
If the coordinates of two points are identical, so are the points themselves.  Hence, we obtain $\mathscr R\tilde\kappa\circ\mathscr R\kappa(q)=\mathscr R\left[\tilde\kappa\kappa\right](q)$,  which is exactly \eqref{grouphomo}. The homomorphism $\mathscr R$ induces a relationship $\mathscr R^*$ between the transpose $\kappa(s)^\mathsf{t}$ of $\kappa(s)$, and the pullback map $\exp(s\xi)^*$, which is defined by $\exp(s\xi)^*=\mathscr R^*\big[\kappa(s)^\mathsf t\big]$. According to \eqref{grouphomo},
\begin{align}
    \notag
    &\mathscr R^*\big[\tilde\kappa(\tilde s)^\mathsf t\big]\circ\mathscr R^*\big[\kappa(s)^\mathsf t\big]=\exp(\tilde s\tilde\xi)^*\circ\exp(s\xi)^*=\big[\exp(s\xi)\circ\exp(\tilde s\tilde\xi)\big]^*\\
&= \big(\mathscr R\kappa(s)\circ\mathscr R\tilde\kappa(\tilde s)\big)^*=\mathscr R\big[\kappa(s)\tilde\kappa(\tilde s)\big]^*=\mathscr R^*\big[\big(\kappa(s)\tilde\kappa(\tilde s)\big)^\mathsf t\big]=\mathscr R^*\big[\tilde\kappa(\tilde s)^\mathsf t\kappa(s)^\mathsf t\big],
\end{align}
 which proves that $\mathscr R^*$ is also well-defined. The homomorphism $\mathscr R^*$ between Lie groups naturally induces a corresponding representation of the Lie algebra, denoted as $\mathsf R$, 
\begin{align}
    \label{inducerepR}
    \xi=\frac{\mathrm d}{\mathrm ds}\exp(s\xi)^*\big|_{s=0}=\frac{\mathrm d}{\mathrm ds}\mathscr R^*\big[\kappa(s)^\mathsf t\big]\big|_{s=0}=\mathsf R\left[\frac{\mathrm d}{\mathrm ds}\kappa(s)^\mathsf t\big|_{s=0}\right]=\mathsf R\big[\kappa^\prime(0)^\mathsf t\big].
\end{align}
Here, the matrix $\kappa^\prime(0)\in\mathfrak{sl}(2,\mathbb R)$ represents the derivative of the matrix $\kappa(s)$ at $s=0$, i.e., at the identity element. Now $\mathfrak{sl}(2,\mathbb R)$ is viewed as the vector space of all $2 \times 2$ real matrices with vanishing trace, since $\det\kappa(s)=1$ implies $\mathrm{tr}\,\kappa^\prime(0)=0$. The pullback map $\exp(s\xi)^*$ acts on functions on the manifold from the left, and consequently acts on points in the manifold from the right, as points and functions are dual to each other. Meanwhile, the matrix $\kappa(s)$, as an element of the Lie group, naturally exhibits a right action. Leveraging the correspondence between the Lie algebra generators and left-invariant vector fields, it is straightforward to compute the Lie bracket. It can be shown that
\begin{align}
    \big[\tilde\xi,\xi\big]=\mathsf R\big[\tilde\kappa^\prime(0)^\mathsf t,\kappa^\prime(0)^\mathsf t\big]=\mathsf R\big[\kappa^\prime(0),\tilde\kappa^\prime(0)\big]^\mathsf t.
\end{align}
That is, the Lie bracket structures of $\kappa^\prime(0)$ and $\xi$ are not identical, but differ by a transpose.

Returning to the analysis of $\kappa(s)$, we will prove that it is real matrix in all cases. If $a^2-4bc>0$, then $\varpi_0, \varpi_1$ form a pair of real roots. It is evident that $\kappa(s)\in\mathrm{SL}(2,\mathbb R)$. Moreover, we can proceed to establish the link between $\kappa(s)$ and $a,b,c$ when $a^2-4bc>0$,
\begin{align}
    \label{kappacoshsinh}
    \notag
    &\kappa(s)=\\
    &\left(\begin{matrix}
        \vspace{12pt}
      \cosh(s\sqrt{a^2-4bc}/2)+\dfrac{a\sinh(s\sqrt{a^2-4bc}/2)}{\sqrt{a^2-4bc}}&\dfrac{2c\sinh(s\sqrt{a^2-4bc}/2)}{\sqrt{a^2-4bc}}\\
      -\dfrac{2b\sinh(s\sqrt{a^2-4bc}/2)}{\sqrt{a^2-4bc}}&\cosh(s\sqrt{a^2-4bc}/2)-\dfrac{a\sinh(s\sqrt{a^2-4bc}/2)}{\sqrt{a^2-4bc}}
    \end{matrix}\right).
\end{align}
We initially did not consider the case where $b = 0$. In this case, if $a\not=0$, it is evident that $a^2-4bc>0$ holds. By directly solving for $t(s)\pm z(s)$, we find that they continue to satisfy eq.\eqref{tskappat}, and the form of $\kappa(s)$ precisely matches the above equation. 

If $a^2-4bc<0$, $\varpi_0, \varpi_1$ constitute a pair of complex conjugates. Nevertheless, by extracting the imaginary factor $\mathrm{i}$ from the matrix elements containing $\sqrt{a^2 - 4bc}$ in \eqref{kappacoshsinh}, and using the identity $\cosh(\sqrt{a^2 - 4bc}) = \cosh(\mathrm{i} \sqrt{4bc - a^2}) = \cos(\sqrt{4bc - a^2})$, along with the similar relationship between the functions $\sinh$ and $\sin$, it is straightforward to transform \eqref{kappacoshsinh} into 
\begin{align}
    \label{kappacossin}
    \notag
    &\kappa(s)=\\
    &\left(\begin{matrix}
    \vspace{12pt}
    \cos(s\sqrt{4bc-a^2}/2)+\dfrac{a\sin(s\sqrt{4bc-a^2}/2)}{\sqrt{4bc-a^2}}&\dfrac{2c\sin(s\sqrt{4bc-a^2}/2)}{\sqrt{4bc-a^2}}\\
    -\dfrac{2b\sin(s\sqrt{4bc-a^2}/2)}{\sqrt{4bc-a^2}}&\cos(s\sqrt{4bc-a^2}/2)-\dfrac{a\sin(s\sqrt{4bc-a^2}/2)}{\sqrt{4bc-a^2}}
\end{matrix}\right).
\end{align}
We observe that all matrix elements of $k(s)$ are also real numbers; hence, $\kappa(s)\in \mathrm{SL}(2,\mathbb R)$. Lastly, when $\varpi_0=\varpi_1$, which is equivalent to $a^2-4bc=0$, we can apply perturbation method, treating $\varpi_0-\varpi_1$ as a small quantity. Therefore, eqs.\eqref{tskappat} and \eqref{SL2Rmatrix} remain valid. To solve for the matrix elements of $\kappa(s)$ in this case, it is sufficient to take the limit $\varpi_1\to\varpi_0$ in eq.\eqref{SL2Rmatrix}, which leads to
\begin{align}
    \label{kappa1}
    \kappa(s)= \left(\begin{matrix}
        \vspace{12pt}
      1-\varpi_0bs &&  \varpi_0^2bs\\
       -bs&&1+\varpi_0bs
    \end{matrix}\right)
    =\left(\begin{matrix}
        \vspace{12pt}
        1+\dfrac{a}{2}s&& cs\\
        -bs&&1-\dfrac{a}{2}s
     \end{matrix}\right).
\end{align}
By directly substituting $\varpi_0=\varpi_1=-\frac{a}{2b}$ into \eqref{tspmzs} instead of using the perturbation method, the above result is reproduced. 
Returning to the case of $b = 0$, which was initially overlooked, and further assuming $a = 0$, by directly solving \eqref{tz}, $t(s)\pm z(s)$ still satisfies \eqref{tskappat}, and $\kappa(s)$ takes the form of the matrix above.

This way, we have demonstrated that the specific form of $\kappa(s)\in\mathrm{SL}(2,\mathbb R)$ generated by any $\xi\in\mathfrak{sl}(2,\mathbb R)$. In particular, for the three special cases $b=c=0$, $a=c=0$, and $a=b=0$, by utilizing eqs.\eqref{kappacoshsinh} and \eqref{kappa1} respectively, we can derive eq.\eqref{oneparatransmation}.

When we replace $t,z$ with $ t^{[i]}(u), z^{[i]}(u)$,  eq.\eqref{tskappat} establishes the relationship between the parameter equations for $\Sigma$ in different coordinate systems. As $ z^{[i]}\sim \epsilon  t^{[i]}$, the transformation of $ t^{[i]}$ can be expressed as
\begin{align}
    \label{SL2Rgroup}
	 t^{[i]}(u)\to \frac{\kappa_{00}\, t^{[i]}(u)+\kappa_{01}}{\kappa_{10}\, t^{[i]}(u)+\kappa_{11}}-\frac{\kappa_{10}( L^{[i]})^2t^{[i]\prime}(u)^2}{\big(\kappa_{10} t^{[i]}(u)+\kappa_{11}\big)^3}\epsilon^2+\mathcal O(\epsilon^4).
\end{align}
It can be observed that unless $\kappa_{10}=0$, which is equivalent to $b=0$ as indicated by eqs.\eqref{kappacoshsinh}\eqref{kappacossin}\eqref{kappa1}, the transformation of $ t^{[i]}(u)$ will involve terms of $\mathcal O(\epsilon^2)$. Conversely, when $b=0$, the strict transformation of $t^{[i]}(u), z^{[i]}(u)$ is given by 
\begin{align}
    \label{expxiaxic}
     t^{[i]}(u)\to \frac{\kappa_{00}}{\kappa_{11}} t^{[i]}(u)+\frac{\kappa_{01}}{\kappa_{11}},
\end{align} 
which ensures that no corrections of $\mathcal O(\epsilon^2)$ are generated. We can now abbreviate the transformation in \eqref{SL2Rgroup} as $\tilde t^{[i]}=\mathscr H[t^{[i]},\epsilon]$. Here, $\mathscr H$ can be interpreted as depending on $\kappa(s),\exp(s\xi)^*$, or $\kappa^\prime(0),\xi$, or the parameters $a, b, c$. The leading-order transformation, namely when $\epsilon \to 0$, is given by
\begin{align}
    \mathscr H[t^{[i]},0]=\frac{\kappa_{00}\, t^{[i]}(u)+\kappa_{01}}{\kappa_{10}\, t^{[i]}(u)+\kappa_{11}}.
\end{align}
Clearly, in the non-trivial case, $\mathscr H[t^{[i]},0]\not=t^{[i]}$. The formal function notation $\mathscr H$ is particularly useful for the conceptual discussion of the iterative expansion of the junction conditions. From \eqref{SL2Rgroup}, it is straightforward to see that 
\begin{align}
    \frac{\partial\mathscr H}{\partial\epsilon}[t^{[i]},0]\equiv 0.
\end{align}

In section \ref{sec3}, we categorize all dilatons in the space $\mathsf{Sol}(\phi)$ into three types based on the sign of $A^2-4BC$. On the other hand, any element $\kappa(s)\in\mathrm{SL}(2,\mathbb R)$ can be classified into hyperbolic, parabolic, and elliptic types\cite{Deitmar2009}, corresponding to $|\mathrm{tr}\,\kappa(s)|>2$, $=2$, and $<2$, respectively. These two classifications are related but not equivalent. We delve into exploring the link between them. The $\mathfrak{sl}(2,\mathbb R)$ is composed of all $2$-order traceless matrices. By employing the commutation relation \eqref{liestructure}, we can establish an isomorphism between $2\times 2$ traceless matrices and Killing vectors,
 \begin{align}
    2\xi_a\longleftrightarrow \left(\begin{matrix}
		1& 0\\
		0&-1
	\end{matrix}\right),\quad
	-\xi_b\longleftrightarrow  \left(\begin{matrix}
		0&1 \\
		0&0
	\end{matrix}\right),\quad
	\xi_c\longleftrightarrow  \left(\begin{matrix}
		0& 0\\
		1&0
	\end{matrix}\right).
 \end{align}
 Through subsequent calculations, as we will see, this isomorphism precisely corresponds to the Lie algebra representation $\mathsf R$ in eq.\eqref{inducerepR}. Then, for any generator $\xi$,
\begin{align}
	\xi=A\xi_a+B\xi_b+C\xi_c=\mathsf R
	\left(\begin{matrix}
        \vspace{9pt}
		      \dfrac{A}{2} && -B           \\
		      C           && -\dfrac{A}{2}
	      \end{matrix}\right)=\mathsf R\big[\kappa^\prime(0)^{\mathsf t}\big].
\end{align}
 It is easy to calculate 
\begin{align}
    \kappa^\prime(0)^2=\frac{1}{4}\big(A^2-4BC\big)\mathbb I,
\end{align}
 where $\mathbb I$ is the unit matrix. With this property, we compute the group element $\kappa(s)=\exp\big(s\kappa^\prime(0)\big)$ using the series expansion of the exponential map. Odd-order terms in the series are proportional to $\kappa^\prime(0)$, while even-order terms are proportional to $\mathbb I$. We obtain
 \begin{align}
	\kappa(s)=
	\begin{cases}
        \vspace{10pt}
		\cosh\big(s\sqrt{A^2-4BC}/2\big)\mathbb I+\dfrac{2\sinh\big(s\sqrt{A^2-4BC}/2\big)}{\sqrt{A^2-4BC}}\kappa^\prime(0),& A^2-4BC>0; \\
        \vspace{10pt}
		\mathbb I+s\kappa^\prime(0) ,                                            & A^2-4BC=0; \\
		\cos\big(s\sqrt{4BC-A^2}/2\big)\mathbb I+\dfrac{2\sin\big(s\sqrt{4BC-A^2}/2\big)}{\sqrt{4BC-A^2}}\kappa^\prime(0),  & A^2-4BC<0.
	\end{cases}
\end{align}
Clearly, upon substituting $A\to a, B\to b, C\to c$, the expression for $\kappa(s)$ precisely matches eqs.\eqref{kappacoshsinh}, \eqref{kappa1}, and \eqref{kappacossin}. This proves that the isomorphism $\mathsf R$ corresponds to the Lie algebra representation in eq.\eqref{inducerepR}.
 
Each dilaton $\phi^{[i]}=A\vartheta_A+B\vartheta_B+C\vartheta_C$ can be associated with a Killing vector $\xi=A\xi_a+B\xi_b+C\xi_c$, and $\xi$ corresponds to a group element $\kappa(1)$ whose trace is
\begin{align}
	\mathrm{tr}\,\kappa(1)=
	\begin{cases}
        \vspace{9pt}
		2\cosh\big(\sqrt{A^2-4BC}/2\big), & A^2-4BC>0; \\
        \vspace{9pt}
		2,                                            & A^2-4BC=0; \\
		2\cos\big(\sqrt{4BC-A^2}/2\big),  & A^2-4BC<0.
	\end{cases}
\end{align}
When $A^2-4BC>0$, there must be $|\mathrm{tr}\,\kappa(1)|>2$, indicating that $\kappa(1)$ corresponds to hyperbolic elements. For $A^2-4BC=0$, $\kappa(1)$ must be parabolic. However, for $A^2-4BC<0$, $\kappa(1)$ does not exclusively correspond to elliptic elements. In fact, when $A^2-4BC=-4\,n^2\pi^2$ with $n$ as any integer, $\kappa(1)=\pm\,\mathbb I$ represents a parabolic element. This phenomenon is evidently rooted in the global topology of $\mathrm{SL}(2,\mathbb R)$. As a topological space, $\mathrm{SL}(2,\mathbb R)$ is not simply connected but homeomorphic to $ \mathbb R^2\times \mathbb S^1$. Arbitrarily choose a $\xi$ with $A^2-4BC<0$, then eq.\eqref{kappacossin} indicates that the curve $\exp(s\xi)$ originating from the unit will necessarily return to the unit for sufficiently large values of $s$. Thus, the integer $n$ reflects the number of windings of the curves $\exp(s\xi)$. Within a sufficiently small open subset of the $\mathsf{Sol}(\phi)$ space, the Lie algebra invariant $A^2-4BC$ and the Lie group invariant $\mathrm{tr}\,\kappa(1)$ are in one-to-one correspondence; thus, classifications based on $A^2-4BC$ and based on $\mathrm{tr}\,\kappa(1)$ are equivalent. In fact, all dilatons satisfying $A^2-4BC\in (-4\pi^2,\infty)$ constitute such a maximal connected open set. Beyond this open set, the overall topology of the Lie group leads to the degeneration of $\mathrm{tr}\,\kappa(1)$.
 
For any element in $\mathrm{SL}(2,\mathbb R)$, it can be represented as the product of three distinct group elements, specifically  $\exp\big(\theta(\xi_b/ L^{[i]}+ L^{[i]}\xi_c)\big)^*\exp(a\xi_a)^*\exp(c\xi_c)^*$. This expression is known as the Iwasawa decomposition~\cite{Deitmar2009}. Significantly, these three group elements precisely correspond to the three residual degrees of freedom associated with the three types of dilaton. Regarding the series expansion of the junction condition and dilaton with respect to $\epsilon$, as per eq.\eqref{expxiaxic}, the transformation $\exp(a\xi_a)^*\exp(c\xi_c)^*$ does not introduce corrections of $\mathcal O(\epsilon^2)$, thus preserving the form of each order in the expansion. Furthermore, interchangeability between the forms of any two different series expansions can be achieved through the action of $\exp\big(\theta(\xi_b/ L^{[i]}+ L^{[i]}\xi_c)\big)^*$ for a certain $\theta$. 

More specifically, we can consider the subgroup consisting of all transformations of the form $\exp(a\xi_a)^*\exp(c\xi_c)^*,~\forall a,c$, denoted as $\mathrm{AN}\subset \mathrm{SL}(2,\mathbb R)$, whose Lie algebra is denoted as $\mathfrak{an}$. The group $\mathrm{AN}$ acts on any $t^{[i]}$ to generate an equivalence class of coordinates, referred to as the $\mathrm{AN}$-class, whose elements all yield the same series expansion. The transformation $\tilde t^{[i]}_\theta\coloneqq\exp\big(\theta(\xi_b/L^{[i]}+ L^{[i]}\xi_c)\big)^*t^{[i]}, ~\forall \theta$, generates all the $\mathrm{AN}$-classes. Therefore, the mapping $\theta\to\tilde t^{[i]}_\theta$ provides a way to label the $\mathrm{AN}$-classes by $\theta$. If $\theta$ undergoes a translation, the labeling remains unchanged in essence. However, this labeling is not unique; it depends on the choice of $t^{[i]}$. For instance, if we choose $\exp(a\xi_a)^*\exp(c\xi_c)^*t^{[i]}$, it is clear that
\begin{align}
    \theta\to \exp\big(\theta(\xi_b/L^{[i]}+ L^{[i]}\xi_c)\big)^*\exp(a\xi_a)^*\exp(c\xi_c)^*t^{[i]}
\end{align}
represents an inequivalent labeling. This is because $\exp\big(\theta(\xi_b/L^{[i]}+ L^{[i]}\xi_c)\big)^*\exp(a\xi_a)^*\exp(c\xi_c)^*t^{[i]}$ and $\exp(\tilde a\xi_a)^*\exp(\tilde c\xi_c)^*\exp\big(\theta(\xi_b/L^{[i]}+ L^{[i]}\xi_c)\big)^*t^{[i]}, \forall \tilde a,\tilde c$, may not belong to the same $\mathrm{AN}$-class. This arises because $\mathrm{AN}$ is not a normal subgroup. To prove this, we need to show that 
\begin{align}
    \label{non-normal}
    \exp\big(\theta(\xi_b/L^{[i]}+ L^{[i]}\xi_c)\big)^*\exp(a\xi_a)^*\exp(c\xi_c)^*\big(-\theta(\xi_b/L^{[i]}+ L^{[i]}\xi_c)\big)^*\not\subset \mathrm{AN},
\end{align}
and we can easily verify that \eqref{non-normal} holds by using
\begin{align}
   \left[\frac{1}{L^{[i]}}\xi_b+ L^{[i]}\xi_c, \xi_a\right]=\frac{1}{L^{[i]}}\xi_b-L^{[i]}\xi_c=\left(\frac{1}{L^{[i]}}\xi_b+L^{[i]}\xi_c\right)-2L^{[i]}\xi_c\not\subset \mathfrak{an}.
\end{align}
If a page $\mathscr V^{[k]}$ carries a Type $-$ dilaton, a natural labeling $\theta\to\tilde t^{[k]}_\theta$ exists on this page. This arises because there is a special coordinate $\tilde t^{[k]}_0=t^{[k]}$ that allows the dilaton to take the form given in \eqref{classifydilaton-}. Furthermore, $\exp\big(\theta(\xi_b/L^{[k]}+ L^{[k]}\xi_c)\big)^*t^{[k]}$ preserves the dilaton's form, meaning that different choices of $\tilde t^{[k]}_0=t^{[k]}$ result only in a translation of the labeling by $\theta$, a feature not shared by the other two types of pages.

\acknowledgments
The author expresses heartfelt gratitude to \textbf{Prof. Cheng Peng} from the Kavli Institute for Theoretical Sciences (KITS), University of Chinese Academy of Sciences, for his indispensable guidance, insightful discussions, and meticulous suggestions throughout the development of this work. His profound physical insights were crucial to the refinement of the theoretical framework and the improvement of the manuscript. 

The author also would like to thank \textbf{Prof. Li-Xin Li} and \textbf{Shaohua Xue} from the Kavli Institute for Astronomy and Astrophysics at Peking University for many stimulating and helpful discussions. 

This work was supported by the \textbf{Scientific Research Startup Fund of Suzhou University of Science and Technology} (Grant No. 332510020).

\bibliographystyle{utcaps}
\bibliography{Refs.bib}

\end{document}